\documentclass[12pt,a4paper]{article}
\usepackage{amsfonts}
\usepackage{mathrsfs}  
\usepackage{epsfig}
\usepackage{axodraw}
\usepackage{graphicx}

\def\simlt{\stackrel{<}{{}_\sim}}

\begin{document}

\title{Third order corrections to the ground state energy of the gas of
  spin $s$ fermions with arbitrary densities of different spin projections}
\author{\em Piotr H. Chankowski, Jacek Wojtkiewicz
  and Szymon Augustynowicz\footnote{Emails: chank@fuw.edu.pl, wjacek@fuw.edu.pl,
    szymon.augustynowicz@student.uw.edu.pl}\\
Faculty of Physics, University of Warsaw,\\
Pasteura 5, 02-093 Warszawa, Poland
}
\maketitle
\abstract{Recently we have computed the third order corrections to the ground
  state energy of the arbitrarily polarized diluted gas of spin 1/2 fermions
  interacting through a spin-independent repulsive two-body potential. Here we
  extend this result to the gas of spin $s$ fermions - a system the Hamiltonian
  of which has an accidental $SU(2s+1)$ symmetry - with arbitrary densities of
  fermions having different spin projections. The corrections are computed
  semi-analytically using the effective field theory approach and are parametrized
  by the $s$- and $p$- wave scattering lengths $a_0$ and $a_1$ and the $s$-wave
  effective radius $r_0$, measurable in the low energy fermion-fermion elastic
  scattering. The result is used to study the impact the higher order corrections
  can have on the characteristics of the phase transition (at zero temperature) to
  the ordered phase (on the emergence of the itinerant ferromagnetism).
\vskip0.1cm

}

\newpage

\section{Introduction}
\label{sec:introd}

Although a clear experimental evidence is still lacking \cite{Jo},
there is a strong conviction that in a finite density system of fermions
a transition to the ordered phase in which the densities of different
spin components are not all equal should be induced by a repulsive
spin-independent interaction, if the system is sufficiently dense and/or
the interaction is strong enough. Such a transition can be conveniently
quantified by a nonzero value of an order parameter $P$ which, in the usually
considered case of spin $s=1/2$ fermions, can be defined to have the meaning
of the polarization (magnetization) $P=(N_\uparrow-N_\downarrow)/N$,
$N=N_\uparrow+N_\downarrow$ of the system. (An analogous order parameter
can be defined in the general case of spin $s>1/2$ if a specific pattern
of the ordering is assumed.)

Theoretical investigation of this transition at zero temperature reduces to
computing the energy density of the system of $N$ fermions enclosed in the
volume $V$ as a function of the densities of different spin projections (or
as a function of the chosen order parameter) and taking the thermodynamic limit.
Such computations are most easily performed using the effective field theory
approach within which the underlying spatially non-local binary interaction of
fermions is replaced by a (in principle infinite) set of local operators of
decreasing length dimension and which yields the expansion of the
computed energy density in powers (in higher orders modified also by
logarithms) of the (overall) Fermi wave vector
\begin{eqnarray}
  k_{\rm F}=\left({6\pi^2\over g_s}~\!{N\over V}\right)^{1/3},\phantom{aaa}
  g_s=2s+1~\!,
\end{eqnarray}
of the system. The expansion obtained in this way is naturally parametrized
by the scattering lengths $a_\ell$, effective radii $r_\ell$, etc.,
$\ell=0,1,\dots,$ which characterize the elastic scattering of low energy
fermions ``in vacuum'' and which are taken to specify the underlying
interaction potential. This approach, pioneered in \cite{HamFur00} (see
also \cite{HamFur02,KolckiSka}), allowed to complete recently \cite{WeDrSch}
the order $k_{\rm F}^4$ (with respect to the energy of the system of completely
non-interacting fermions) corrections to the ground state of the system of spin
$s$ fermions with equal densities of different spin projections. Using this
approach it was also easy to reproduce (semi-analytically) but in the universal
setting the old (obtained by considering the specific model of hard spheres
interaction) result of Kanno
\cite{KANNO} who computed (fully analytically) the order $k_{\rm F}^2$ correction
to the ground state of the system of spin $1/2$ fermions \cite{CHWO1} for an
arbitrary value of the system's polarization $P$. The effective theory approach
allows to immediately extend \cite{PECABO} the result of Kanno to the case of
spin $s$ fermions:
the complete up to the order $k_{\rm F}^2$ formula for the ground state energy
density $E_\Omega/V$ as a function of the densities of fermions having
different spin projections can be now written in the form\footnote{The order
  $k_{\rm F}$ correction is the extension (obvious from the effective field
  theory point of view) of the textbook
  \cite{Kesio,Pathria} mean-field result to the case of spin $s$ fermions.}
\begin{eqnarray}
  {E_\Omega\over V}={k_{\rm F}^3\over6\pi^2}~\!{3\over5}~\!
  {\hbar^2k_{\rm F}^2\over2m_f}\left\{\sum_{\sigma=1}^{g_s} x_\sigma^5
  +{20\over9\pi}~\!k_{\rm F}a_0\sum_{\sigma^\prime<\sigma}x_{\sigma^\prime}^3x_\sigma^3 
  +{2\over\pi^2}~\!(k_{\rm F}a_0)^2\sum_{\sigma^\prime<\sigma}
  J_K(x_{\sigma^\prime},x_\sigma,)\right\},\label{eqn:EnergyUpToSecondOrder}
\end{eqnarray}
in which 
\begin{eqnarray}
  x_\sigma={p_{{\rm F}\sigma}\over k_{\rm F}}~\!,\phantom{aaaa}
  p_{{\rm F}\sigma}=\left({6\pi^2}~\!{N_\sigma\over V}\right)^{1/3},\label{eqn:Defs}
\end{eqnarray}
and 
\begin{eqnarray}
  J_K(x,y)={1\over21}\left\{
  (22~\! x^3y^3 (x + y) - 4~\!x ^7 \ln\!\left(1+{y\over x}\right) - 
  4~\!y^7 \ln\!\left(1+{x\over y}\right)\right.\nonumber\\
  + {1\over2}~\!x~\!y~\!(x - y)^2 (x + y) [15~\!(x^2 + y^2) + 11~\! x~\! y]
  \phantom{aaaaa}~\nonumber\\
   \left.- {7\over4} (x - y)^4 (x + y) ((x + y)^2 + x~\!y)
      \ln\!\left({x+y\over x-y}\right)\right\},\phantom{a}~\label{eqn:JKanno}
\end{eqnarray}
is the function obtained by Kanno in his computation \cite{KANNO}.
The extension to $s>1/2$ is of obvious interest in view of the fact that
the systems of interacting fermions (and bosons) are nowadays investigated
in experiments with cold atoms in which the interaction strength
can be tuned by exploiting the physics of the so-called Feshbach resonance.

The system of spin of $1/2$ fermions with the (spin independent) repulsive
interaction constitutes the simple, textbook example \cite{Kesio,Pathria} of
the continuous phase transition because such a character of the paramagnetic
to ferromagnetic transition to the ordered state (signaled by the emergence
of spontaneous magnetization), both when it is induced by decreasing
the temperature and by increasing the density and/or the interaction strengths
(that is when the value of the parameter $k_{\rm F}a_0$ is increased) is found
when the computations are restricted to the mean field approximation (only the
order $k_{\rm F}a_0$ corrections to the thermodynamic functions are taken into
account).
However when the order $(k_{\rm F}a_0)^2$ corrections are included
this transition turns at low temperatures into the ordinary first order one as
has been found in \cite{DUMacDO} and at exactly $T=0$ can be easily confirmed
by using the formula (\ref{eqn:EnergyUpToSecondOrder}) \cite{CHWO2,PECABO}.
Although in this approximation the transition at zero temperature occurs at
$k_{\rm F}a_0=1.054$ at which value the
perturbative expansion cannot most probably be trusted,
such a character of the transition seemed to be in line with the arguments
\cite{BeKiVoj} based on general principles of the Ginzburg-Landau theory. 

On the other hand in the series of papers \cite{He} (overlooked in
\cite{CHWO1,CHWO2}) the ground state energy of the system of spin $1/2$
fermions as a function of the polarization $P$ has been computed beyond
a fixed order, by performing a resummation of an infinite subclass of the
effective field theory diagrams contributing to it. As the authors of these
papers claim, the transition to the ordered state at $T=0$ becomes then
again of the continuous type. Moreover, the critical value of the parameter
$k_{\rm F}a_0$ found by them is in a surprisingly good agreement with the
estimates based on the Quantum Monte Carlo computations reported in
\cite{QMC10}. While assessing the reliability of the approximations made in
\cite{He} is in general difficult, it might be instructive to see, using the
complete order $k_{\rm F}^3$ corrections, which we have recently computed in
\cite{CHWO3} in the case of spin $1/2$, how the third order terms
included in the resummation compare with the ones neglected and how the 
character of the phase transition changes when it is analyzed 
using the complete third order formula for the ground state energy. In view
of the mentioned circumstance that in
experimental setups the role of fermions are played by cold atomic gases 
which can have spins greater than $1/2$, it is also of interest to investigate
the transition in the case of higher spins. Finally, according to the power
counting rules which organize the expansion within the effective field theory
\cite{Weinberg,HamFur00}, in the order $k_{\rm F}^3$ to the energy contribute 
density for the first time interaction operators involving in addition to four
fermion field operators also two spatial derivatives which bring in the
dependence on the $p$-wave scattering length $a_1$ and the $s$-wave effective
radius $r_0$ which parametrize the departure of the result from strict
unversality. It is therefore natural to check their potential impact on the
phase transition. 

Therefore the plan of the paper is as follows. In the next section, to
prepare the ground, we recall the effective theory approach allowing to
easily recover the order $k_{\rm F}a_0$ and $(k_{\rm F}a_0)^2$ (this one
semi-analytically) corrections in (\ref{eqn:EnergyUpToSecondOrder}).
Then, in Section \ref{sec:Third} and \ref{sec:HigherDimOps} we determine
the next order correction to the formula (\ref{eqn:EnergyUpToSecondOrder})
adapting appropriately the computation \cite{CHWO3} of the third order
correction to the ground state energy of the arbitrarily polarized gas of
spin $1/2$ fermions. The discussion of the approach of \cite{He}
is included in Section \ref{sec:Third}. The character of the transition
to the ordered state is discussed in Section \ref{sec:PhTr}. 


\section{Second order corrections}

As said, the effective theory naturally yields an expansion of the computed
quantities in powers of the product $kR$ of the wave vector $k$ corresponding
to the characteristic energy scale of the problem and of the characteristic
length scale $R$ of the ``fundamental'' interaction potential. In the case of
the corrections to the ground state energy of the gas of fermions the role of
the characteristic wave vector plays the Fermi wave vector $k_{\rm F}$ defined
in (\ref{eqn:Defs}) and they are most conveniently computed (using the standard
Feynman rules formulated e.g. in \cite{FetWal}) as the sum of connected
``vacuum'' diagrams using the Dyson expansion of the right hand side of the
formula
\begin{eqnarray}
  \lim_{T\rightarrow\infty}\exp(-iT(E_\Omega-E_{\Omega_0})/\hbar)=
  \lim_{T\rightarrow\infty}\langle\Omega_0
  |{\rm T}\exp\!\left(-{i\over\hbar}\!\int_{-T/2}^{T/2}\!dt~\!
  V^I_{\rm int}(t)\right)\!|\Omega_0\rangle~\!,\label{eqn:basicFormula}
\end{eqnarray}
in which $|\Omega_0\rangle$ is the ground state of the free Hamiltonian in
the $N$-fermion subspace of the Fock space, $V^I_{\rm int}(t)$ is the theory
interaction operator $V_{\rm int}$ taken in the interaction picture and the
limits $T\rightarrow\infty$, $V\rightarrow\infty$ are implicit (T stands
for the chronological product.

If, as assumed, the underlying ''fundamental'' interaction of spin $s$
fermions is spin-independent and (in the infinite volume limit) Galileo
invariant, $V_{\rm int}$ in (\ref{eqn:basicFormula}) is the effective theory
interaction operator\footnote{The absence of the $\sigma^\prime=\sigma$
term in the sum is the immediate consequence of the anticommutativity of the
fermionic field operators; in \cite{CHWO1}, where only the spin $s=1/2$ was
treated, the sum in (\ref{eqn:Veff}) consisted a single term only.}
of the form \cite{HamFur00}
\begin{eqnarray} 
V_{\rm int}=C_0\!\int\!d^3\mathbf{x}
\sum_{\sigma^\prime<\sigma}\psi_{\sigma^\prime}^\dagger\psi_{\sigma^\prime}
\psi_\sigma^\dagger\psi_\sigma
+V_{\rm int}^{(C_2)}+V_{\rm int}^{(C_2^\prime)}+\dots,  \label{eqn:Veff}
\end{eqnarray}
The terms $V_{\rm int}^{(C_2)}$ and
$V_{\rm int}^{(C_2^\prime)}$, which will be specified in Section
\ref{sec:HigherDimOps}, involve two spatial derivatives and are of lower
length dimension than the first one; the ellipsis stands for other
operators of yet lower length dimension.
Despite the infinite number of interaction terms in (\ref{eqn:Veff}),
to a term proportional to a given power\footnote{Terms proportional to
  higher powers of $(kR)$ are also modified by logarithms \cite{HamFur00}.}
of $(kR)$ in the computed quantity contributes only a finite number of
Feynman diagrams (constructed using only a finite subset of the interaction
terms of (\ref{eqn:Veff})) - only those dictated by the so-called power
counting rules \cite{Weinberg,HamFur00}.
The couplings $C_0$, $C_2$, $C_2^\prime,\dots$ of the effective theory
are fixed by matching the amplitude of the elastic fermion-fermion
scattering computed using the effective theory onto the general form
of such an amplitude expanded in powers of the relative particle momenta
and parametrized by the scattering lengths $a_\ell$, the effective radii $r_\ell$,
etc. which with the sufficient accuracy characterize  the ``fundamental''
binary interaction. Of course, since the effective interaction is local,
divergences appear and must be regularized (in \cite{CHWO1} a cutoff
$\Lambda$ on the wave vectors circulating in loops have been imposed); the
relations between computed physical quantities like $a_\ell$,
$r_\ell$, $E_\Omega,\dots$ and the effective theory parameters $C_0$, $C_2,\dots$
depend then on the regularization but this dependence disappears when the
computed energy is expressed in terms of the scattering lengths and radii.
To the order which is needed to compute the third order corrections to
the ground state energy of the gas of fermions
the relations between the couplings $C_0$, $C_2$ and $C_2^\prime$
and the scattering lengths $a_0$, $a_1$ and the radius $r_0$ are known
\cite{WeDrSch} and (when the cutoff $\Lambda$ is used as the regulator) read
\begin{eqnarray}
  C_0(\Lambda)={4\pi\hbar^2\over m_f}~\!a_0\left(1
  +{2\over\pi}~\!a_0\Lambda+{4\over\pi^2}~\!a^2_0\Lambda^2+\dots\right),
 \phantom{aaaaaaa}\! \label{eqn:C0Determined}\\
 C_2(\Lambda)={4\pi\hbar^2\over m_f}~\!{1\over2}~\!a_0^2r_0+\dots~\!,
 \phantom{aaaa}
 C_2^\prime(\Lambda) ={4\pi\hbar^2\over m_f}~\!a_1^3+\dots~\!.
 \label{eqn:C2C2primeDetermined}
\end{eqnarray}

\begin{figure}[]
\begin{center}
\begin{picture}(370,40)(5,0)
\ArrowArc(30,20)(25,70,290)
\DashArrowArc(30,20)(25,290,70){2}
\DashArrowArc(50,20)(25,110,250){2}
\ArrowArc(50,20)(25,250,110)
\Vertex(40,-2.5){2}
\Vertex(40,42.5){2}
\Text(155,30)[]{$A_{\sigma^\prime,\sigma}(q)=$}
\Vertex(190,30){2}
\ArrowArc(210,35)(20,195,345)
\DashArrowArcn(210,25)(20,165,15){2}
\Vertex(230,30){2}
\Text(215,50)[]{$^{-k+q~\sigma^\prime}$}
\Text(215,8)[]{$_{k~\sigma}$}
\Text(295,30)[]{$B_{\sigma^\prime,\sigma}(q)=$}
\Vertex(330,30){2}
\ArrowArc(350,35)(20,195,345)
\DashArrowArc(350,25)(20,15,165){2}
\Vertex(370,30){2}
\Text(355,50)[]{$^{k~\sigma^\prime}$}
\Text(355,8)[]{$_{k+q~\sigma}$}
\end{picture}
\end{center}
\caption{The only type of diagrams contributing to the order $(k_{\rm F}R)^2$
  correction to the ground state energy of the gas of spin $s$ fermions and
  two ``elementary'' one-loop diagrams out of which the second
  order and the third order corrections with the $C_0$ couplings can be
  constructed. Solid and dashed lines denote propagators of fermions
  with different spin projections.}
\label{fig:ElementaryLoops}
\end{figure}
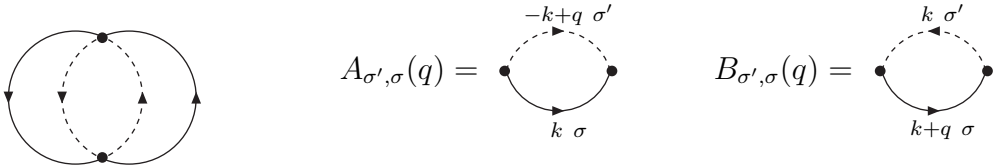

The power counting rules tell that the order $k_{\rm F}R$ and $(k_{\rm F}R)^2$
corrections to the ground state energy of the gas of fermions that
is, the order $k_{\rm F}a_0$ and $(k_{\rm F}a_0)^2$ terms in the formula
(\ref{eqn:EnergyUpToSecondOrder}), are generated solely by the term in
(\ref{eqn:Veff}) proportional to $C_0$. The correction of order $k_{\rm F}a_0$
is therefore given by the sum over all possible choices of the (different)
pairs of spin projections $\sigma^\prime$ and $\sigma$ circulating in the
two loops of a simple two-loop diagram that can be formed by sewing
together the corresponding pairs of lines representing the
interaction vertex proportional to $C_0$ while that of order $(k_{\rm F}a_0)^2$
are given by the similar sum over spin projections circulating
in the three-loop diagram shown in Figure \ref{fig:ElementaryLoops} on the
left. This diagram can be composed in three equivalent ways (corresponding
to different assignments of the momenta to its internal lines):
\begin{eqnarray}
  {1\over i\hbar}~\!{E^{(2)}_\Omega\over V}
  ={1\over2!}\left({C_0\over i\hbar}\right)^2\sum_{\sigma^\prime<\sigma}
  \int\!{d^4q\over(2\pi)^4}~\![A_{\sigma^\prime,\sigma}(q)]^2
  ={1\over2!}\left({C_0\over i\hbar}\right)^2\sum_{\sigma^\prime<\sigma}
  \int\!{d^4q\over(2\pi)^4}~\![B_{\sigma^\prime,\sigma}(q)]^2\nonumber\\
  ={1\over2!}\left({C_0\over i\hbar}\right)^2\sum_{\sigma^\prime<\sigma}
  \int\!{d^4q\over(2\pi)^4}~\!B_{\sigma,\sigma}(q)~\!
  B_{\sigma^\prime,\sigma^\prime}(q)~\!,\phantom{aaaaaaaaaaaaaaaaa}
  \label{eqn:ProductsOfTwoBlocks}
\end{eqnarray}
out of the two ``elementary blocks'' (corresponding to the
two ``elementary'' loops shown in Figure \ref{fig:ElementaryLoops})
\begin{eqnarray}
A_{\sigma^\prime,\sigma}(q)=\int\!{d^3\mathbf{k}\over(2\pi)^3}~\!i\!
\left[{\theta(|\mathbf{k}-\mathbf{q}|-p_{{\rm F}\sigma^\prime})~\!
    \theta(|\mathbf{k}|-p_{{\rm F}\sigma})\over
q^0-\omega_{\mathbf{k}}-\omega_{\mathbf{k}-\mathbf{q}}+i0}
  -{\theta(p_{{\rm F}\sigma^\prime}-|\mathbf{k}-\mathbf{q}|)~\!
    \theta(p_{{\rm F}\sigma}-|\mathbf{k}|)\over
    q^0-\omega_{\mathbf{k}}-\omega_{\mathbf{k}-\mathbf{q}}-i0}\right],
\phantom{a}\nonumber\\
\label{eqn:blockA}\\
B_{\sigma^\prime,\sigma}(q)=\int\!{d^3\mathbf{k}\over(2\pi)^3}~\!i\!
\left[-{\theta(p_{{\rm F}\sigma^\prime}-|\mathbf{k}|)~\!
    \theta(|\mathbf{k}+\mathbf{q}|-p_{{\rm F}\sigma})\over
q^0+\omega_{\mathbf{k}}-\omega_{\mathbf{k}+\mathbf{q}}+i0}
  +{\theta(|\mathbf{k}|-p_{{\rm F}\sigma^\prime})~\!
    \theta(p_{{\rm F}\sigma}-|\mathbf{k}+\mathbf{q}|)\over
q^0+\omega_{\mathbf{k}}-\omega_{\mathbf{k}+\mathbf{q}}-i0}\right],\nonumber\\
\label{eqn:blockB}
\end{eqnarray}
in which $\omega_{\mathbf{k}}=\hbar\mathbf{k}^2/2m_f$, obtained by the standard
integration over $k^0$ using the residue method. Each of these ``blocks'' is a
sum of two parts - those in the ``elementary block'' $A_{\sigma^\prime,\sigma}(q)$
which will be needed in the discussion of the approach of \cite{He}, can be
given the interpretation as corresponding to the propagation of two particles
(the first part) and the propagation of two holes (the other part). In either
form of (\ref{eqn:ProductsOfTwoBlocks}) after the integration over $dq^0$ only
two (out of the possible four) terms are nonzero (and equal to one another),
because in the two others all poles are on the same side of the real $q^0$ axis.

The final form of the order $(k_{\rm F}a_0)^2$ contribution to the energy
density is obtained (see \cite{CHWO1}) by making the appropriate change
of the remaining integration variables. In this way one arrives at
\begin{eqnarray}
  {E^{(2)}_\Omega\over V}={64m_fC_0^2\over\hbar^2(2\pi)^6}\sum_{\sigma^\prime<\sigma}
  J(p_{{\rm F}\sigma^\prime},~\!p_{{\rm F}\sigma})~\!,
\end{eqnarray}
with the function $J(p_{{\rm F}\sigma},~\!p_{{\rm F}\sigma^\prime})
=J(p_{{\rm F}\sigma^\prime},~\!p_{{\rm F}\sigma})$ given by the integral 
\begin{eqnarray}
  J(p_{{\rm F}\sigma^\prime},~\!p_{{\rm F}\sigma})
  =\int_0^{s_{\rm max}}\!ds~\!s^2~\!{1\over4\pi}\!\int\!d^3\mathbf{t}~\!
  \theta(p_{{\rm F}\sigma^\prime}-|\mathbf{u}+\mathbf{s}|)~\!
  \theta(p_{{\rm F}\sigma}-|\mathbf{t}-\mathbf{s}|)~\!
  g_{\sigma^\prime,\sigma}(t,s)~\!,\label{eqn:JfunctionDef}
\end{eqnarray}
in which $s_{\rm max}={1\over2}(p_{{\rm F}\sigma^\prime}+p_{{\rm F}\sigma})$.
The function $g_{\sigma^\prime,\sigma}(t,s)\equiv
g_{\sigma^\prime,\sigma}(|\mathbf{t}|,~\!|\mathbf{s}|)=g_{\sigma,\sigma^\prime}(t,s)$
given by the integral (symmetric in the labels $\sigma$ and $\sigma^\prime$)
\begin{eqnarray}
g_{\sigma^\prime,\sigma}(t,s)\equiv{1\over4\pi}\!\int\!d^3\mathbf{u}~\!
{\theta(|\mathbf{u}+\mathbf{s}|-p_{{\rm F}\sigma^\prime})
\theta(|\mathbf{u}-\mathbf{s}|-p_{{\rm F}\sigma})
\over \mathbf{t}^2-\mathbf{u}^2+i0}\nonumber\\
=-\Lambda+g^{\rm fin}_{\sigma^\prime,\sigma}(t,s)+{t^2\over\Lambda}+\dots,
\phantom{aaaaaaaaaaaa}\label{eqn:g(t,s)functionDef}
\end{eqnarray}
has been obtained in \cite{CHWO1} in the analytic form which assumes (without
loosing generality) that $p_{{\rm F}\sigma^\prime}\leq p_{{\rm F}\sigma}$:
\begin{eqnarray}
  g^{\rm fin}_{\sigma^\prime,\sigma}(t,s)={1\over2}~\!p_{{\rm F}\sigma}
  +{t\over4}\ln{(p_{{\rm F}\sigma}-t)^2-s^2\over(p_{{\rm F}\sigma}+t)^2-s^2}
  +{p^2_{{\rm F}\sigma}-s^2-t^2\over8s}\ln{(p_{{\rm F}\sigma}+s)^2
    -t^2\over(p_{{\rm F}\sigma}-s)^2-t^2}~\!,\label{eqn:g(t,s)Explicit1}
\end{eqnarray}
when $0<s\leq{1\over2}(p_{{\rm F}\sigma}-p_{{\rm F}\sigma^\prime})$ and
\begin{eqnarray}
  g^{\rm fin}_{\sigma^\prime,\sigma}(t,s)={1\over4}(p_{{\rm F}\sigma^\prime}+p_{{\rm F}\sigma}
  +2s)+{t\over4}\ln{p_{{\rm F}\sigma^\prime}+s-t\over p_{{\rm F}\sigma^\prime}+s+t}
  +{t\over4}\ln{p_{{\rm F}\sigma}+s-t\over p_{{\rm F}\sigma}+s+t}\phantom{aaaaaaaaa}
  \nonumber\\
  +~\!{p^2_{{\rm F}\sigma^\prime}-t^2-s^2\over8s}
  \ln{(p_{{\rm F}\sigma^\prime}+s)^2-t^2\over u_0^2-t^2}
  +{p^2_{{\rm F}\sigma}-t^2-s^2\over8s}
  \ln{(p_{{\rm F}\sigma}+s)^2-t^2\over u_0^2-t^2}~\!,\label{eqn:g(t,s)Explicit2}
\end{eqnarray}
where
\begin{eqnarray}
  u_0^2={1\over2}(p_{{\rm F}\sigma^\prime}^2+p_{{\rm F}\sigma}^2)-s^2~\!,
  \label{eqn:u0squared}
\end{eqnarray}
when ${1\over2}(p_{{\rm F}\sigma}-p_{{\rm F}\sigma^\prime})<s\leq s_{\rm max}$.

It is convenient to write the function
$J(p_{{\rm F}\sigma^\prime},~\!p_{{\rm F}\sigma})$ as the sum
\begin{eqnarray}
  J=J_{\rm div}+J_{\rm fin}+J_{1/\Lambda}+\dots,\nonumber
\end{eqnarray}
where the successive terms directly correspond to the terms in
(\ref{eqn:g(t,s)functionDef}). $J_{\rm fin}$ is therefore independent of the
cutoff $\Lambda$ while $J_{1/\Lambda}$ and the ellipsis stand for terms vanishing
in the limit $\Lambda\rightarrow\infty$. As
\begin{eqnarray}
  \int_0^{s_{\rm max}}\!ds~\!s^2~\!{1\over4\pi}\!
  \int\!d^3\mathbf{t}~\!
    \theta(p_{{\rm F}\sigma^\prime}-|\mathbf{u}+\mathbf{s}|)~\!
    \theta(p_{{\rm F}\sigma}-|\mathbf{t}-\mathbf{s}|)
    ={p^3_{{\rm F}\sigma^\prime}p^3_{{\rm F}\sigma}\over72}~\!,\label{eqn:stIntegrals}
\end{eqnarray}
\cite{CHWO1}, $J_{\rm div}=-(\Lambda/72)p^3_{{\rm F}\sigma^\prime}p^3_{{\rm F}\sigma}$
and it is straightforward to check that the divergences arising from
$J_{\rm div}$ cancel against similar terms proportional to $\Lambda$ arising
when $C_0$ in the correction 
\begin{eqnarray}
  {E^{(1)}_\Omega\over V}={C_0\over36\pi^4}\sum_{\sigma^\prime<\sigma}
  p^3_{{\rm F}\sigma^\prime}p^3_{{\rm F}\sigma}~\!,\label{eqn:C0Correction}
\end{eqnarray}
arising from the mentioned two-loop diagrams is expressed in terms of the
$s$-wave scattering length $a_0$ as in (\ref{eqn:C0Determined}). The cutoff
independent terms of (\ref{eqn:C0Correction}) give then the term proportional
to $k_{\rm F}a_0$ in (\ref{eqn:EnergyUpToSecondOrder})
and the terms proportional to inverse powers of $\Lambda$ (which would be
absent, if the dimensional regularization was used instead of the cutoff
$\Lambda$) can be, if the computation is restricted to the second order,
simply discarded. In the computation including higher order corrections they
cancel against similar spurious contributions arising in higher orders.

If the densities of fermions having different spin projections are all
equal ($N_\sigma=N/g_s$, $p_{{\rm F}\sigma}=k_{\rm F}$ for all $\sigma=1,\dots,g_s$),
one numerically finds that
\begin{eqnarray}
  J_{\rm fin}(k_{\rm F},k_{\rm F})={k_{\rm F}^7\over840}~\!(11-2\ln2)~\!.\nonumber
\end{eqnarray}
The sum over pairs of different spin projections gives in this case the factor
${1\over2}g_s(g_s-1)$ and one recovers the well known old result re-derived in
\cite{HamFur00} using the effective field theory. If the densities of
fermions with different spin projections are not equal, numerical evaluation
of the integrals defining the function
$J(p_{{\rm F}\sigma^\prime},~\!p_{{\rm F}\sigma})$ shows it is proportional to the
function $J_K$ (\ref{eqn:JKanno}) introduced by Kanno the precise relation
being\footnote{From the formulae (\ref{eqn:JfunctionDef}) and
  (\ref{eqn:g(t,s)functionDef}) it is clear that
  $J(p_{{\rm F}\sigma^\prime},~\!p_{{\rm F}\sigma})\equiv
  J(x_{\sigma^\prime}k_{\rm F},~\!x_\sigma k_{\rm F})=
  k_{\rm F}^7J(x_{\sigma^\prime},~\!x_{\sigma})$.}
\begin{eqnarray}
  160~\!J_{\rm fin}(p_{{\rm F}\sigma^\prime},~\!p_{{\rm F}\sigma})
  =k_{\rm F}^7J_K(x_{\sigma^\prime},~\!x_\sigma)~\!.\nonumber
\end{eqnarray}
This leads to the result (\ref{eqn:EnergyUpToSecondOrder})
obtained recently in a somewhat different way in \cite{PECABO}.

\section{Third order corrections proportional to $C_0^3$}
\label{sec:Third}

According to the power counting rules the order
$(k_{\rm F}R)^3$ corrections to the ground state energy density arise from
the four-loop diagrams with three interaction vertices proportional to $C_0$
and from the two-loop diagrams with a single $C_2$ or $C_2^\prime$ interaction
vertex (the terms $V_{\rm int}^{(C_2)}$ and $V_{\rm int}^{(C_2^\prime)}$ in
(\ref{eqn:Veff})). In this section we work out the contribution of the
first class of diagrams; the contribution of diagrams with the vertices
generated by the interaction terms proportional to $C_2$ and $C_2^\prime$
will be computed in Section \ref{sec:HigherDimOps}.

\begin{figure}[]
\begin{center}
\begin{picture}(360,80)(5,0)
  \ArrowArc(40,40)(40,330,90)
  \ArrowArc(40,40)(40,90,210)
  \ArrowArc(40,40)(40,180,360)
  \Vertex(74,20){2}
  \Vertex(6,20){2}
  \Vertex(40,80){2}
  \DashArrowArcn(40,-53)(80,115,65){2}
  \DashArrowArcn(120,86)(80,235,185){2}
  \DashArrowArcn(-40,86)(80,355,305){2}
  \ArrowArc(200,40)(40,330,90)
  \ArrowArc(200,40)(40,90,210)
  \ArrowArc(200,40)(40,180,360)
  \Vertex(234,20){2}
  \Vertex(166,20){2}
  \Vertex(200,80){2}
  \DashArrowArc(200,-53)(80,65,115){2}
  \DashArrowArc(280,86)(80,185,235){2}
  \DashArrowArc(120,86)(80,305,355){2}
  \DashArrowArc(320,40)(40,330,90){3}
  \DashArrowArc(320,40)(40,90,210){1}
  \ArrowArc(320,40)(40,210,330)
  \Vertex(354,20){2}
  \Vertex(286,20){2}
  \Vertex(320,80){2}
  \ArrowArc(320,-53)(80,65,115)
  \DashArrowArc(400,86)(80,185,235){3}
  \DashArrowArc(240,86)(80,305,355){1}
%
\end{picture}
\end{center}
\caption{The single particle-particle and two different particle-hole
  vacuum diagrams contributing in the order $C^3_0$ to $E_\Omega$. Different
  types of lines represent propagators of fermions having
  different spin projections. The second particle-hole diagram exists
only if there are more than two different spin projections ($s>{1\over2}$).}
\label{fig:C0cube}
\end{figure}
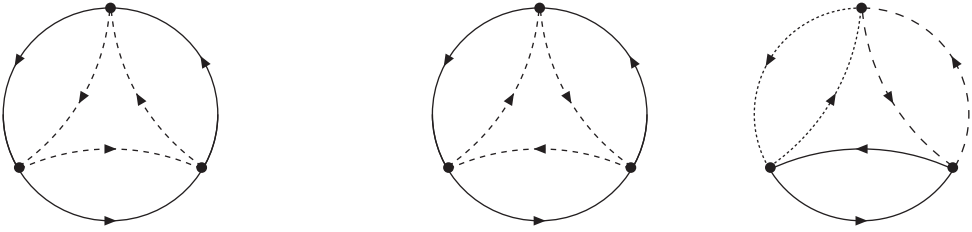

If $s>{1\over2}$ three kinds of nonvanishing four-loop diagrams with three
interaction vertices proportional to the $C_0$ coupling which can be formed.
They are shown in Figure \ref{fig:C0cube}. The first one is the so-called
particle-particle diagram while the remaining two are called particle-hole
diagrams.

The contribution of the order $C_0^3$ particle-particle diagram shown in
Figure \ref{fig:C0cube} on the left is (2 is the combinatoric factor)
\begin{eqnarray}
  {1\over i\hbar}~\!{E^{(3){\rm pp}}_\Omega\over V}
  ={1\over3!}\left({C_0\over i\hbar}\right)^3 2\sum_{\sigma^\prime<\sigma}
  \int\!{d^4q\over(2\pi)^4}~\![A_{\sigma^\prime,\sigma}(q)]^3~\!,
  \label{eqn:ThreeAblocksComposed}
\end{eqnarray}
As it is clear from the spin structure of the interaction term
proportional to $C_0$ and the flow of the spin projections in the
diagram, the sum runs only over pairs of different spin projection
labels $\sigma^\prime\neq\sigma$. There are $g_s(g_s-1)/2$ such pairs and,
if densities of fermions with different spin projections are all equal,
the summation reproduces the usual spin factor associated with
this contribution \cite{HamFur00}. For the sake of further discussion
we note that the product of three $A$-''blocks'' gives rise to
$2^3=8$ terms out of two are killed by the integration over $dq^0$
(they have all poles on the same side of the real $q^0$ axis)
while the remaining six split into three identical terms which arise
from the products of two terms corresponding in (\ref{eqn:blockA})
- appealing to the interpretation of the two terms of the $A$-''block'' -
to the propagation of ``particles'' and one corresponding to the
propagation of holes and another three identical ones in which
this composition is reversed.
As a result this contribution to the energy density can be written
in terms of two functions, $G^{(1)}(p_{{\rm F}\sigma^\prime},p_{{\rm F}\sigma})$,
and $G^{(2)}(p_{{\rm F}\sigma^\prime},p_{{\rm F}\sigma})$ (corresponding
respectively to these two kinds of compositions): 
\begin{eqnarray}
  {E^{(3){\rm pp}}_\Omega\over V}
  ={128m_f^2C_0^3\over(2\pi)^8\hbar^4}\sum_{\sigma^\prime<\sigma}
  \left[G^{(1)}(p_{{\rm F}\sigma^\prime},p_{{\rm F}\sigma})
    +G^{(2)}(p_{{\rm F}\sigma^\prime},p_{{\rm F}\sigma})
    \right].
\end{eqnarray}
The functions $G^{(1)}(p_{{\rm F}\sigma^\prime},p_{{\rm F}\sigma})$ and
$G^{(2)}(p_{{\rm F}\sigma^\prime},p_{{\rm F}\sigma})$ are given by the integrals 
\begin{eqnarray}
 G^{(1)}(p_{{\rm F}\sigma^\prime},p_{{\rm F}\sigma})=\!\int_0^{s_{\rm max}}\!\!ds~\!s^2~\!
  {1\over4\pi}\!\int\!d^3\mathbf{t}~\!
  \theta(p_{{\rm F}\sigma^\prime}-|\mathbf{t}+\mathbf{s}|)~\!
  \theta(p_{{\rm F}\sigma}-|\mathbf{t}-\mathbf{s}|)~\!
  (g_{\sigma^\prime,\sigma}(t,s))^2~\!,\phantom{a}\!\label{eqn:G1FunctionDef}
\end{eqnarray}
\begin{eqnarray}
 G^{(2)}(p_{{\rm F}\sigma^\prime},p_{{\rm F}\sigma})=\!\int_0^{s_{\rm max}}\!\!ds~\!s^2~\!
  {1\over4\pi}\!\int\!d^3\mathbf{t}~\!
  \theta(|\mathbf{t}+\mathbf{s}|-p_{{\rm F}\sigma^\prime})~\!
  \theta(|\mathbf{t}-\mathbf{s}|-p_{{\rm F}\sigma})~\!
  (h_{\sigma^\prime,\sigma}(t,s))^2~\!,\phantom{a}\!\label{eqn:G2FunctionDef}
\end{eqnarray}
in which $g_{\sigma^\prime,\sigma}(t,s)$ is the function defined by
(\ref{eqn:g(t,s)functionDef}) and
$h_{\sigma^\prime,\sigma}(t,s)\equiv h_{\sigma^\prime,\sigma}(|\mathbf{t}|,|\mathbf{s}|)
=h_{\sigma,\sigma^\prime}(t,s)$ is given by the 
finite integral
\begin{eqnarray}
  h_{\sigma^\prime,\sigma}(t,s)={1\over4\pi}\!\int\!d^3\mathbf{u}~\!
  {\theta(p_{{\rm F}\sigma^\prime}-|\mathbf{u}+\mathbf{s}|)
    \theta(p_{{\rm F}\sigma}-|\mathbf{u}-\mathbf{s}|)\over
    \mathbf{t}^2-\mathbf{u}^2-i0}~\!.\label{eqn:h(t,s)functionDef}
\end{eqnarray}
The explicit form of the function $h_{\sigma^\prime,\sigma}(t,s)$ has been
obtained\footnote{Analysis of the integration domains of
  (\ref{eqn:h(t,s)functionDef}) shows that the pole at
$\mathbf{t}^2=\mathbf{u}^2$ is never reached.}
using the same method (based on the trick introduced in Appendix C of
\cite{FUHATI}) by which the function $g_{\sigma^\prime,\sigma}(t,s)$ has been
computed in \cite{CHWO1}. Assuming without loss of generality
that $p_{{\rm F}\sigma^\prime}\leq p_{{\rm F}\sigma}$, it reads
\begin{eqnarray}
  h_{\sigma^\prime,\sigma}(t,s)=-{1\over2}~\!p_{{\rm F}\sigma^\prime}
  -{t\over4}\ln{t-(p_{{\rm F}\sigma^\prime}-s)\over t+(p_{{\rm F}\sigma^\prime}-s)}
  -{t\over4}\ln{t-(p_{{\rm F}\sigma^\prime}+s)\over t+(p_{{\rm F}\sigma^\prime}+s)}
  \phantom{aaaaaaaa}\nonumber\\
  +~\!{t^2-(p_{{\rm F}\sigma^\prime}^2-s^2)\over8s}
  \ln{t^2-(p_{{\rm F}\sigma^\prime}+s)^2\over t^2-(p_{{\rm F}\sigma^\prime}-s)^2}~\!,
  \phantom{aaaaaaaaaaaaaaaaa}~ \label{eqn:h(t,s)Explicit:slts0}
\end{eqnarray}
if $0\leq s\leq {1\over2}(p_{{\rm F}\sigma}-p_{{\rm F}\sigma^\prime})$, and ($u^2_0$
is again given by (\ref{eqn:u0squared}))
\begin{eqnarray}
  h_{\sigma^\prime,\sigma}(t,s)
  ={1\over2}\left(2s-p_{{\rm F}\sigma^\prime}-p_{{\rm F}\sigma}\right)
  -{t\over4}\ln{t-(p_{{\rm F}\sigma^\prime}-s)\over t+(p_{{\rm F}\sigma^\prime}-s)}
  -{t\over4}\ln{t-(p_{{\rm F}\sigma}-s)\over t+(p_{{\rm F}\sigma}-s)}
  \phantom{aaaaaa}\nonumber\\
    -{1\over8s}\left[(p_{{\rm F}\sigma}-s)^2+(p_{{\rm F}\sigma^\prime}-s)^2-2u_0^2\right]
    \phantom{aaaaaaaaaaaaaaaaaaaaaaaaaaa}\label{eqn:h(t,s)Explicit:s0gts}\\
    -~\!{t^2-p_{{\rm F}\sigma}^2+s^2\over8s}
    \ln{t^2-(p_{{\rm F}\sigma}-s)^2\over t^2-u_0^2}
   -{t^2-p_{{\rm F}\sigma^\prime}^2+s^2\over8s}
   \ln{t^2-(p_{{\rm F}\sigma^\prime}-s)^2\over t^2-u_0^2}~\!,\nonumber
\end{eqnarray}
if ${1\over2}(p_{{\rm F}\sigma}-p_{{\rm F}\sigma^\prime})\leq s\leq s_{\rm max}
={1\over2}(p_{{\rm F}\sigma}+p_{{\rm F}\sigma^\prime})$. Expanding  both these
expression one finds that the function $h_{\sigma^\prime,\sigma}(t,s)$ behaves
as $1/t^2$ when $t=|\mathbf{t}|\rightarrow\infty$. Therefore, although the
integral over $d^3\mathbf{t}$ in (\ref{eqn:G2FunctionDef}) covers the
infinite exterior of the two Fermi spheres, it is convergent - unlike the
function $G^{(1)}(p_{{\rm F}\sigma^\prime},p_{{\rm F}\sigma})$ which depends on the
divergent function $g_{\sigma^\prime,\sigma}(t,s)$, 
the function $G^{(2)}(p_{{\rm F}\sigma^\prime},p_{{\rm F}\sigma})$ is finite.
\vskip0.1cm

Since the remaining order $C_0^3$ contributions to $E_\Omega/V$ will turn out
finite, already at this point one can check the cancellation of ultraviolet
divergences in the expression
\begin{eqnarray}
  {\Delta E_\Omega\over V}=\sum_{\sigma^\prime<\sigma}\left[
    {p_{{\rm F}\sigma^\prime}^3p_{{\rm F}\sigma}^3\over36\pi^4}~\!C_0
  +{64 m_fC_0^2\over(2\pi)^6\hbar^2}~\!J(p_{{\rm F}\sigma^\prime},p_{{\rm F}\sigma})
  +{128m_f^2C_0^3\over(2\pi)^8\hbar^4}~\!
  G^{(1)}(p_{{\rm F}\sigma^\prime},p_{{\rm F}\sigma})\right]+\dots\nonumber
\end{eqnarray}
Inserting here the expansions (\ref{eqn:C0Determined}) of $C_0$ (truncated
to the appropriate order), writing
\begin{eqnarray}
  (g_{\sigma^\prime,\sigma}(t,s))^2=\Lambda^2-2\Lambda~\! g^{\rm fin}_{\sigma^\prime,\sigma}
  +(g^{\rm fin}_{\sigma^\prime,\sigma})^2-2t^2+\dots\label{eqn:g(t,s)Squared}
\end{eqnarray}
and using result (\ref{eqn:stIntegrals}) it is a matter of a simple
algebra to check that all terms diverging with $\Lambda\rightarrow\infty$
as well as the spurious terms specific for the adopted regularization
(that is the finite term arising from the product of the term proportional to
$a_0\Lambda$ and the term $t^2/\Lambda$ from $g_{\sigma^\prime,\sigma}(t,s)$
in the contribution proportional to $C_0^2$ and the term $-2t^2$
in (\ref{eqn:g(t,s)Squared})) completely cancel out as they should.
As a result, to obtain the relevant contribution to $E_\Omega/V$
one can simply drop all $\Lambda$ dependent terms in the
function $g_{\sigma^\prime,\sigma}(t,s)$ and set $C_0=(4\pi\hbar^2/m_f)a_0$ (the
remaining part of the function $G^{(1)}$ will be denoted $G^{(1)}_{\rm fin}$).
\vskip0.2cm

The first type of the particle-hole diagrams (the middle one in Figure
\ref{fig:C0cube}) has been computed in \cite{CHWO3} in the case of spin
$s={1\over2}$ fermions. The generalization of the formulae obtained
there is straightforward and reduces to taking again the sum over pairs of
different spin projections:
\begin{eqnarray}
  {1\over i\hbar}~\!{E^{(3){\rm ph1}}_\Omega\over V}
  ={1\over3!}\left({C_0\over i\hbar}\right)^3 2\sum_{\sigma^\prime<\sigma}
  \int\!{d^4q\over(2\pi)^4}~\![B_{\sigma^\prime,\sigma}(q)]^3~\!.
\end{eqnarray}
(Again  2 is the combinatoric factor). This gives
\begin{eqnarray}
  {E_\Omega^{(3){\rm ph1}}\over V}=-{32m_f^2C_0^3\over(2\pi)^8\hbar^4}
  \sum_{\sigma^\prime<\sigma}
  \left[K^{(1)}(p_{{\rm F}\sigma^\prime},p_{{\rm F}\sigma})
    +K^{(2)}(p_{{\rm F}\sigma^\prime},p_{{\rm F}\sigma})\right],
  \label{eqn:C03phEnergyPol1}
\end{eqnarray}
where the functions $K^{(1)}(p_{{\rm F}\sigma^\prime},p_{{\rm F}\sigma})$ and
$K^{(2)}(p_{{\rm F}\sigma^\prime},p_{{\rm F}\sigma})$ are given by
\begin{eqnarray}
  K^{(1)}(p_{{\rm F}\sigma^\prime},p_{{\rm F}\sigma})=\int_0^\infty\!ds~\!s^2~\!{1\over4\pi}\!
  \int\!d^3\mathbf{t}~\!\theta(|\mathbf{t}+\mathbf{s}|-p_{{\rm F}\sigma^\prime})~\!
  \theta(p_{{\rm F}\sigma}-|\mathbf{t}-\mathbf{s}|)~\!
  (f_{\sigma^\prime,\sigma}^{(1)}(\mathbf{t}\!\cdot\!\mathbf{s},~\!s))^2~\!,\phantom{a}
  \label{eqn:K1functionDef}
\end{eqnarray}
\begin{eqnarray}
  K^{(2)}(p_{{\rm F}\sigma^\prime},p_{{\rm F}\sigma})=\int_0^\infty\!ds~\!s^2~\!{1\over4\pi}\!
  \int\!d^3\mathbf{t}~\!\theta(p_{{\rm F}\sigma^\prime}-|\mathbf{t}+\mathbf{s}|)~\!
  \theta(|\mathbf{t}-\mathbf{s}|-p_{{\rm F}\sigma})~\!
  (f_{\sigma^\prime,\sigma}^{(2)}(\mathbf{t}\!\cdot\!\mathbf{s},~\!s))^2~\!,
  \phantom{a}\label{eqn:K2functionDef}
\end{eqnarray}
and the functions
$f_{\sigma^\prime,\sigma}^{(1)}(\mathbf{t}\!\cdot\!\mathbf{s},~\!s)$ and
$f_{\sigma^\prime,\sigma}^{(2)}(\mathbf{t}\!\cdot\!\mathbf{s},~\!s)$ are defined
by the integrals
\begin{eqnarray}
f_{\sigma^\prime,\sigma}^{(1)}(\mathbf{t}\!\cdot\!\mathbf{s},~\!s)
={1\over4\pi}\!\int\!d^3\mathbf{u}~\!
{\theta(p_{{\rm F}\sigma^\prime}-|\mathbf{u}+\mathbf{s}|)~\!
\theta(|\mathbf{u}-\mathbf{s}|-p_{{\rm F}\sigma})\over(\mathbf{u}-\mathbf{t})\cdot
\mathbf{s}+i0}~\!,\label{eqn:f1(t,s)Def}
\end{eqnarray}
and 
\begin{eqnarray}
f_{\sigma^\prime,\sigma}^{(2)}(\mathbf{t}\!\cdot\!\mathbf{s},~\!s)
={1\over4\pi}\!\int\!d^3\mathbf{u}~\!
{\theta(|\mathbf{u}+\mathbf{s}|-p_{{\rm F}\sigma^\prime})~\!
\theta(p_{{\rm F}\sigma}-|\mathbf{u}-\mathbf{s}|)\over(\mathbf{u}-\mathbf{t})\cdot
\mathbf{s}-i0}~\!.\label{eqn:f2(t,s)Def}
\end{eqnarray}

Finally the contribution of the last type of diagrams (the rightmost
one in Figure \ref{fig:C0cube}) is given by
\begin{eqnarray}
{1\over i\hbar}~\!{E^{(3){\rm ph2}}_\Omega\over V}
={1\over3!}\left({C_0\over i\hbar}\right)^3(-3!)\!
\sum_{\sigma^{\prime\prime},\sigma^\prime,\sigma}
\int\!{d^4q\over(2\pi)^4}~\!B_{\sigma^{\prime\prime},\sigma^{\prime\prime}}(q)~\!
B_{\sigma^\prime,\sigma^\prime}(q)~\!B_{\sigma,\sigma}(q)~\!,
\end{eqnarray}
where $-3!$ is the sign/combinatoric factor and the sum is over all triplets
of different spin projections ($\sigma\neq\sigma^\prime\neq\sigma^{\prime\prime}$,
$\sigma\neq\sigma^{\prime\prime}$). After the standard steps this leads to
\begin{eqnarray}
{E_\Omega^{(3){\rm ph2}}\over V}={32m_f^2C_0^3\over(2\pi)^8\hbar^4}
\sum_{\sigma^{\prime\prime},\sigma^\prime,\sigma}
\left\{
  \tilde K^{(1)}(p_{{\rm F}\sigma^{\prime\prime}};~\!p_{{\rm F}\sigma^\prime},p_{{\rm F}\sigma})
 +\tilde K^{(2)}(p_{{\rm F}\sigma^{\prime\prime}};~\!p_{{\rm F}\sigma^\prime},p_{{\rm F}\sigma})
  \right.\phantom{a}~\nonumber\\
 +\tilde K^{(1)}(p_{{\rm F}\sigma^\prime};~\!p_{{\rm F}\sigma},p_{{\rm F}\sigma^{\prime\prime}})
 +\tilde K^{(2)}(p_{{\rm F}\sigma^\prime};~\!p_{{\rm F}\sigma},p_{{\rm F}\sigma^{\prime\prime}})
\phantom{aa}\label{eqn:C03phEnergyPol2}\\
\left.
 +\tilde K^{(1)}(p_{{\rm F}\sigma};~\!p_{{\rm F}\sigma^{\prime\prime}},p_{{\rm F}\sigma^\prime})
 +\tilde K^{(2)}(p_{{\rm F}\sigma};~\!p_{{\rm F}\sigma^{\prime\prime}},p_{{\rm F}\sigma^\prime})
\right\},\nonumber
\end{eqnarray}
where the functions
$\tilde K^{(1)}(p_{{\rm F}\sigma^{\prime\prime}};~\!p_{{\rm F}\sigma^\prime},p_{{\rm F}\sigma})$,
and
$\tilde K^{(2)}(p_{{\rm F}\sigma^{\prime\prime}};~\!p_{{\rm F}\sigma^\prime},p_{{\rm F}\sigma})$,
symmetric in their second and third Fermi momentum labels, are given by the
following nested integrals
\begin{eqnarray}
  \tilde K^{(1)}(p_{{\rm F}\sigma^{\prime\prime}};~\!p_{{\rm F}\sigma^\prime},p_{{\rm F}\sigma})
  \phantom{aaaaaaaaaaaaaaaaaaaaaaaaaaaaaaaaaaaaaaaaaaaaaaaaaaaa}\nonumber\\
  =\int_0^\infty\!ds~\!s^2~\!{1\over4\pi}\!
  \int\!d^3\mathbf{t}~\!
  \theta(|\mathbf{t}+\mathbf{s}|-p_{{\rm F}\sigma^{\prime\prime}})~\!
  \theta(p_{{\rm F}\sigma^{\prime\prime}}-|\mathbf{t}-\mathbf{s}|)~\!
  f_{\sigma^\prime,\sigma^\prime}^{(1)}(\mathbf{t}\!\cdot\!\mathbf{s},~\!s)~\!
  f_{\sigma,\sigma}^{(1)}(\mathbf{t}\!\cdot\!\mathbf{s},~\!s)~\!,\phantom{a}
  \label{eqn:tildeK1functionDef}
\end{eqnarray}
\begin{eqnarray}
  \tilde K^{(2)}(p_{{\rm F}\sigma^{\prime\prime}};~\!p_{{\rm F}\sigma^\prime},p_{{\rm F}\sigma})
  \phantom{aaaaaaaaaaaaaaaaaaaaaaaaaaaaaaaaaaaaaaaaaaaaaaaaaaaa}\nonumber\\
  =\int_0^\infty\!ds~\!s^2~\!{1\over4\pi}\!
  \int\!d^3\mathbf{t}~\!
  \theta(p_{{\rm F}\sigma^{\prime\prime}}-|\mathbf{t}+\mathbf{s}|)~\!
  \theta(|\mathbf{t}-\mathbf{s}|-p_{{\rm F}\sigma^{\prime\prime}})~\!
  f_{\sigma^\prime,\sigma^\prime}^{(2)}(\mathbf{t}\!\cdot\!\mathbf{s},~\!s)~\!
  f_{\sigma,\sigma}^{(2)}(\mathbf{t}\!\cdot\!\mathbf{s},~\!s)~\!,\phantom{a}
  \label{eqn:tildeK2functionDef}
\end{eqnarray}
However, the change $\mathbf{s}\rightarrow-\mathbf{s}$ in the integrals
defining the functions $\tilde K^{(1)}$ and $\tilde K^{(2)}$ makes it obvious that
\begin{eqnarray}
\tilde K^{(2)}(p_{{\rm F}\sigma^{\prime\prime}};~\!p_{{\rm F}\sigma^\prime},p_{{\rm F}\sigma})
=\tilde K^{(1)}(p_{{\rm F}\sigma^{\prime\prime}};~\!
p_{{\rm F}\sigma^\prime},p_{{\rm F}\sigma})~\!.
\end{eqnarray}
This allows to simplify the right hand side of (\ref{eqn:C03phEnergyPol2}) by
retaining in it (and doubling) only, say, the functions $\tilde K^{(2)}$.

If the densities $N_\sigma/V$ of fermions having different spin projection are
all equal
($p_{{\rm F}\sigma}=p_{{\rm F}\sigma^\prime}=p_{{\rm F}\sigma^{\prime\prime}}=k_{\rm F}$), then
$\tilde K^{(2)}(k_{\rm F};~\!k_{\rm F},k_{\rm F})=K^{(1)}(k_{\rm F},k_{\rm F})
=K^{(2)}(k_{\rm F},k_{\rm F})$. The spin factors of the sum of the contributions
(\ref{eqn:C03phEnergyPol1}) and (\ref{eqn:C03phEnergyPol2}) combine in this
case into
\begin{eqnarray}
  -{g_s(g_s-1)\over2}+3~\!{g_s(g_s-1)(g_s-2)\over6}
  ={1\over2}~\!g_s(g_s-1)(g_s-3)~\!,\nonumber
\end{eqnarray}
which is the spin factor associated with the particle-hole diagram
in this case \cite{HamFur00}.

Analysis of the integration domains in the formulae defining the
$K$-functions one finds that the poles of the integrands at
$\mathbf{u}\cdot\mathbf{s}=\mathbf{t}\cdot\mathbf{s}$ are never reached,
so the prescriptions $\pm i0$ can be dropped. Still, obtaining the analytic
forms of the functions
$f_{\sigma^\prime,\sigma}^{(1,2)}(\mathbf{t}\!\cdot\!\mathbf{s},~\!s)$ is rather
cumbersome - hence we do not give the technical details of the computation.
The final formulae (different in the regimes of small, intermediate and
large $s$) for these functions can be found in \cite{CHWO3}; the formulae
for the function $f^{(2)}_{\sigma,\sigma}(\mathbf{s}\cdot\mathbf{t},s)$ (not
given in \cite{CHWO3}) can in principle be obtained as a
$p_{{\rm F}\sigma^\prime}=p_{{\rm F}\sigma}$ limit
of the function $f^{(2)}_{\sigma^\prime,\sigma}(\mathbf{s}\cdot\mathbf{t},s)$
but for this particular case a much simpler formula can be
obtained.\footnote{We have, of course,
  checked numerically that
  $f^{(2)}_{\sigma^\prime,\sigma}(\mathbf{s}\cdot\mathbf{t},s)$
  agrees with $f^{(2)}_{\sigma,\sigma}(\mathbf{s}\cdot\mathbf{t},s)$
  in the limit $p_{{\rm F}\sigma^\prime}=p_{{\rm F}\sigma}$.}
It reads:
\begin{eqnarray}
f^{(2)}_{\sigma\sigma}(\mathbf{t}\cdot\mathbf{s},s)=
  s~\!(p_{{\rm F}\sigma}-t)
  \phantom{aaaaaaaaaaaaaaaaaaaaaaaaaaaaaaaaaaaaaaaaaaaaa}\nonumber\\
  +{1\over4}(p_{{\rm F}\sigma}^2-s^2+2st)
  \ln\!\left[{(t^2-p_{{\rm F}\sigma}^2+s^2-2st)
      (t p_{{\rm F}\sigma}-p_{{\rm F}\sigma}^2+s^2-st)\over t^2(
    t p_{{\rm F}\sigma}+k^2-s^2+st)}\right]\nonumber\\
    -{1\over4}(p_{{\rm F}\sigma}^2-s^2-2st)
    \ln\!\left[{(t^2-p_{{\rm F}\sigma}^2+s^2+2st)(tp_{{\rm F}\sigma}
        -p_{{\rm F}\sigma}^2+s^2+st)\over t^2
      (t p_{{\rm F}\sigma}+p_{{\rm F}\sigma}^2-s^2-st)}\right]\nonumber\\
      +{1\over2}(p_{{\rm F}\sigma}^2-s^2)
      \ln{p_{{\rm F}\sigma}+s\over p_{{\rm F}\sigma}-s}
      -t^2\ln{t-s-p_{{\rm F}\sigma}\over t+s-p_{{\rm F}\sigma}}~\!
      \phantom{aaaaaaaaaaaaaaaaaaaa}\nonumber
      \end{eqnarray}
$f^{(2)}_{\sigma\sigma}(\mathbf{t}\cdot\mathbf{s},s)
=-f^{(1)}_{\sigma\sigma}(\mathbf{t}\cdot\mathbf{s},s)$.

With the analytical formulae for these functions the remaining integrals in
(\ref{eqn:K1functionDef}), (\ref{eqn:K2functionDef})
and (\ref{eqn:tildeK2functionDef}) can be evaluated numerically using
the standard Mathematica built-in routine for multidimensional integration
over a prescribed domain. 
\vskip0.1cm

The resulting, order $(k_{\rm F}a_0)^3$ correction, that adds up to the terms in
the curly bracket in (\ref{eqn:EnergyUpToSecondOrder}) is\footnote{From the
  defining formulae it is clear that all the functions
  $G^{(1,2)}(p_{{\rm F}\sigma^\prime},p_{{\rm F}\sigma})$,
  $K^{(1,2)}(p_{{\rm F}\sigma^\prime},p_{{\rm F}\sigma})$ and
  $\tilde K^{2)}(p_{{\rm F}\sigma^{\prime\prime}};p_{{\rm F}\sigma^\prime},p_{{\rm F}\sigma})$ 
  scale like the eight power of the wave vector $k_{\rm F}$ which effectively
  reduces them to functions of the single variable $r=x_{\sigma^\prime}/x_\sigma$:
  $G^{(1)}(p_{{\rm F}\sigma^\prime},p_{{\rm F}\sigma})
  =p_{{\rm F}\sigma}^8G^{(1)}(x_{\sigma^\prime}/x_\sigma,1)$, etc.}
  \begin{eqnarray}
  {160\over\pi^3}~\!(k_{\rm F}a_0)^3\left(\sum_{\sigma^\prime<\sigma}\!
    \left[4G^{(1)}_{\rm fin}(x_{\sigma^\prime},x_\sigma)+4G^{(2)}(x_{\sigma^\prime},x_\sigma)
    -K^{(1)}(x_{\sigma^\prime},x_\sigma)-K^{(2)}(x_{\sigma^\prime},x_\sigma)\right]\right.
    \nonumber\\
    \left.+\sum_{\sigma^{\prime\prime},\sigma^\prime,\sigma}\!\!
    \left[2\tilde K^{(2)}(x_{\sigma^{\prime\prime}};x_{\sigma^\prime},x_\sigma)
      +2\tilde K^{(2)}(x_{\sigma^\prime};x_\sigma,x_{\sigma^{\prime\prime}})
      +2\tilde K^{(2)}(x_\sigma;x_{\sigma^{\prime\prime}},x_{\sigma^\prime})\right]\right),
    \phantom{a}\label{eqn:finalC0cubeContribution}
\end{eqnarray}
where the sum in the last line is over all triplets of different
indices $\sigma^{\prime\prime},\sigma^\prime,\sigma$.
\vskip0.1cm

\begin{figure}
\psfig{figure=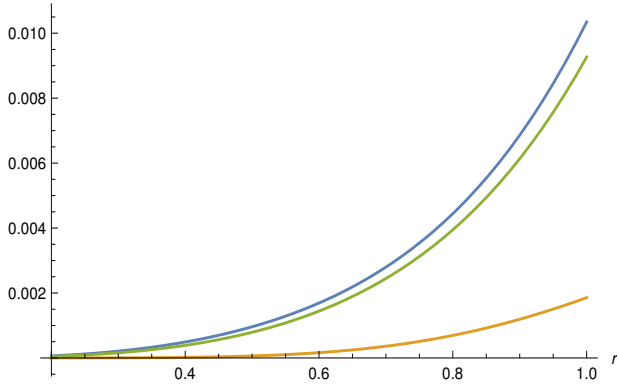,width=8.cm,height=5.0cm} 
\caption{Plot of the functions $G^{(1)}_{\rm fin}(r,1)$ (the highest, blue, curve),
  $G^{(2)}(r,1)$ (the lowest, red, curve) and
  ${1\over4}[K^{(1)}(r,1)+K^{(2)}(r,1)]$ (the middle, green, curve).}
\label{fig:G1G2Kplot}
\end{figure}

With these results one can try to verify the reliability of the approach
developed in \cite{He}. Its authors have considered only the case of spin
$1/2$ fermions and only the contributions to the ground state energy density
which, like the order $C_0^2$ contribution (\ref{eqn:ProductsOfTwoBlocks})
and the order $C_0^3$ contribution (\ref{eqn:ThreeAblocksComposed}), can be
composed out of the $A$-''blocks''. Of the corresponding $n$-th order
contribution ($(n-1)!$ is the combinatoric factor)
\begin{eqnarray}
  {1\over i\hbar}~\!{E^{(n)\rm pp}_\Omega\over V}
  ={1\over n!}\left({C_0\over i\hbar}\right)^n(n-1)!\sum_{\sigma^\prime<\sigma}
  \!\int\!{d^4q\over(2\pi)^4}~\![A_{\sigma^\prime,\sigma}(q)]^n~\!,
  \label{eqn:AtoNthContribution}
\end{eqnarray}
which gives rise to $2^n-2$ nonvanishing terms they retain only $n$
identical ones obtained by taking $n-1$ parts of the $A$-''block''
corresponding to the propagation of particles and one part corresponding
to the propagation of holes (again appealing to the interpretation
introduced earlier). Thus, out of the complete order $C_0^3$ contribution
obtained above they take only the function
$G^{(1)}_{\rm fin}(p_{{\rm F}\sigma^\prime},p_{{\rm F}\sigma})$ discarding the
contributions of the function $G^{(2)}(p_{{\rm F}\sigma^\prime},p_{{\rm F}\sigma})$
and of the sum ${1\over4}[K^{(1)}(p_{{\rm F}\sigma^\prime},p_{{\rm F}\sigma})
  +K^{(2)}(p_{{\rm F}\sigma^\prime},p_{{\rm F}\sigma})]$. As the plot
\ref{fig:G1G2Kplot} shows, the first of these two functions is indeed one
order of magnitude smaller than $G^{(1)}_{\rm fin}$ in the entire range of the
ratio $r=p_{{\rm F}\sigma^\prime}/p_{{\rm F}\sigma}$ (both these functions, similarly
as the functions $K^{(1,2)}$ and $\tilde K^{(2)}$ vanish at $r=0$ as a result
of the absence of $s$-wave interaction of fermions obeying the Pauli
exclusion principle) but the same is not true of the second
function which is always comparable with $G^{(1)}_{\rm fin}$ and enters 
(\ref{eqn:finalC0cubeContribution}) with the opposite sign strongly
reducing the contribution of the function $G^{(1)}_{\rm fin}$. Moreover, in higher
orders the number of discarded terms arising from (\ref{eqn:AtoNthContribution})
rapidly grows and if each of the discarded $(2^n-2)-n$ terms of the contribution
(\ref{eqn:AtoNthContribution}) is only one order of magnitude smaller than
each of the $n$ terms taken into account, the domination of the latter can
be invalidated (as the transition to the ordered state is expected to occur at
$k_{\rm F}a_0\simlt1$, the higher order contributions are not much suppressed by
the powers of the expansion parameter $k_{\rm F}a_0$). Moreover, in
higher orders the number of other types of diagrams (with topologies different
than those shown in Figure \ref{fig:C0cube}) grows and their contribution,
like the one of $K^{(1)}(p_{{\rm F}\sigma^\prime},p_{{\rm F}\sigma})
+K^{(2)}(p_{{\rm F}\sigma^\prime},p_{{\rm F}\sigma})$, may dominate the
one of the retained terms. Thus the reliability of the
approximations made in \cite{He} may be questioned even though
the numerical results obtained there agree quite well with those
obtained by the Quantum Monte Carlo methods.

\section{Contribution of the interaction terms proportional to $C_2$
  and $C_2^\prime$}
\label{sec:HigherDimOps}

According to the power counting rule in the order
$k^3_{\rm F}(\hbar^2k_{\rm F}^2/2m_f)(k_{\rm F}R)^3$ one has to take
into account also  the operators:
\begin{eqnarray}
V^{(C_2)}_{\rm int}=-{C_2\over8}\int\!d^3\mathbf{x}
\sum_{\sigma<\sigma^\prime}\left[\psi^\dagger_\sigma\psi^\dagger_{\sigma^\prime}
(\psi_{\sigma^\prime}\mbox{\boldmath{$\nabla$}}^2\psi_\sigma
-2\mbox{\boldmath{$\nabla$}}\psi_{\sigma^\prime}\!\cdot\!
\mbox{\boldmath{$\nabla$}}\psi_\sigma
+\mbox{\boldmath{$\nabla$}}^2\psi_{\sigma^\prime}~\!\psi_\sigma)\right.
\phantom{a}~\nonumber\\
\left.+(\mbox{\boldmath{$\nabla$}}^2\psi^\dagger_\sigma~\!\psi^\dagger_{\sigma^\prime}
-2\mbox{\boldmath{$\nabla$}}\psi^\dagger_\sigma\!\cdot\!
\mbox{\boldmath{$\nabla$}}\psi^\dagger_{\sigma^\prime}
+\psi^\dagger_\sigma\mbox{\boldmath{$\nabla$}}^2\psi^\dagger_{\sigma^\prime})
\psi_{\sigma^\prime}\psi_\sigma\right],\label{eqn:C2Interaction}
\end{eqnarray}
and
\begin{eqnarray}
  V_{\rm int}^{(C_2^\prime)}= -{C^\prime_2\over8}\!\int\!d^3\mathbf{x}
  \sum_{\sigma,{\sigma^\prime}}
(\mbox{\boldmath{$\nabla$}}\psi^\dagger_\sigma\psi^\dagger_{\sigma^\prime}
-\psi^\dagger_\sigma\mbox{\boldmath{$\nabla$}}\psi^\dagger_{\sigma^\prime})\!\cdot\!
(\mbox{\boldmath{$\nabla$}}\psi_{\sigma^\prime}\psi_\sigma-
\psi_{\sigma^\prime}{\mbox{\boldmath{$\nabla$}}}\psi_\sigma)~\!.
\label{eqn:C2PrimOpExplicit}
\end{eqnarray}
(Notice that in the second operator the sum involves also terms with
$\sigma={\sigma^\prime}$).
Their contribution to the ground state energy density in the case of
nonequal densities of spin $1/2$ fermions having different spin projections
has been explicitly written down in \cite{CHWO3}. Generalization of these
formulae to the case of spin $s$ fermions with different densities of their
$g_s$ spin projections is straightforward. They read
\begin{eqnarray}
  {E^{(C_2)}_\Omega\over V}={C_2\over240\pi^4}\sum_{\sigma^\prime<\sigma}
  p_{{\rm F}\sigma^\prime}^3p_{{\rm F}\sigma}^3(p_{{\rm F}\sigma^\prime}^2+p_{{\rm F}\sigma}^2)
  \phantom{aaaaaaaaaaaaaaaaaaaaaa}~\nonumber\\
  ={k_{\rm F}^3\over6\pi^2}~\!{3\over5}~\!{\hbar^2k_{\rm F}^2\over2m_f}~\!
  {1\over6\pi}~\!k_{\rm F}^3a_0^2r_0
  \sum_{\sigma^\prime<\sigma} x^3_{\sigma^\prime}x^3_\sigma(x^2_{\sigma^\prime}+x^2_\sigma )~\!.
  \phantom{aaaaaaaaaaaa}
\end{eqnarray}
\begin{eqnarray}
  {E^{(C^\prime_2)}_\Omega\over V}={C^\prime_2\over120\pi^4}\left\{
  \sum_{\sigma=1}^{g_s}p_{{\rm F}\sigma}^8+{1\over2}\sum_{\sigma^\prime<\sigma}
  p_{{\rm F}\sigma^\prime}^3p_{{\rm F}\sigma}^3(p_{{\rm F}\sigma^\prime}^2
  +p_{{\rm F}\sigma}^2)\right\}\phantom{aaaaaaaaaa}\nonumber\\
  ={k_{\rm F}^3\over6\pi^2}~\!{3\over5}~\!{\hbar^2k_{\rm F}^2\over2m_f}~\!
  {2\over3\pi}~\!(k_{\rm F}a_1)^3
  \left\{\sum_{\sigma=1}^{g_s}x_\sigma^8+{1\over2}\sum_{\sigma^\prime<\sigma}
  x^3_{\sigma^\prime}x^3_\sigma(x^2_{\sigma^\prime}+x^2_\sigma)\right\}.
\end{eqnarray}
For equal all densities of all spin projections this reduces to the known
results given e.g. in \cite{HamFur00}. The contribution $E^{(C^\prime_2)}_\Omega$
is special in that unlike the remaining ones (to this order) it does not
vanish if all fermions have the same spin projection. This has important
consequences for the transition to the ordered state.

\section{Phase transition to the ordered state}
\label{sec:PhTr}

A short range repulsive interaction, cooperating with the Pauli exclusion rule
can, if its strength is large enough and the gas density is not too low, cause
the transition to the ordered state. The qualitative explanation of this
phenomenon is that because of the Pauli exclusion, the dominant $s$-wave
interaction energy of two fermions with the same spin projection must vanish
(it is positive if the spin projections are different); hence assuming by the
majority of fermions the same spin projection and decreasing thereby the
interaction energy may become energetically more favourable than minimizing
the kinetic energy by having all spin projections equally populated. Since the
effect is due to the competition between the energy of the free system and the
interaction energy, it can occur only if the interaction is sufficiently 
strong, which implies that it hardly can be treated perturbatively.
In the mean field approximation,
equivalent to taking in (\ref{eqn:EnergyUpToSecondOrder}) into account only the
order $k_{\rm F}a_0$ correction, and for spin $s=1/2$ fermions this happens when
$k_{\rm F}a_0>\pi/2$ (see e.g. \cite{Stoner,Kesio,Pathria}). 

In this simplest case the role of the order parameter naturally plays the
polarization $P=(N_+-N_-)/N$ and seeking the minimum of the ground state energy
$E_\Omega(P)$ as a function of $P$ is straightforward. In the case of spin $s$
fermions there are in principle many possible configurations with non-equal
densities of different spin projections but from the
heuristic argument given above it follows that in this case the minimum of
the ground state energy should be in the configuration in which only one spin
projection has the density larger than the remaining ones which all have
equal densities.\footnote{The authors of \cite{PECABO}  claim to
  have proved this mathematically but a cursory look at their argument
  which relies on the obvious symmetry of the equations determining the
densities of the spin projections different than the distinguished one
shows it is not valid.}
Of course, in the ordered phase there are then $g_s=2s+1$ degenerate ground
states related to one another by the $SU(g_s)$ symmetry because in the
absence of an external field any of the $g_s$ spin projections can be that one
which is populated by the majority of fermions. Therefore the natural order
parameter $P$ defined by setting (without loss of generality we assume that
it is the first spin component which is populated differently than the
remaining ones)
\begin{eqnarray}
  x_1=(1+P)^{1/3}~\!,\phantom{aaa}
  x_\sigma=\left(1-{P\over g_s-1}\right)^{1/3}~\!,\phantom{aaa}
  \sigma=2,\dots,g_s~\!,\label{eqn:x1Andx2Defs}\\
  x_1^3+x_2^3+\dots+x_{g_s}^3=g_s~\!,\phantom{aaaa}
  -1\leq P\leq g_s-1~\!,\phantom{aaaaaaa}
\end{eqnarray}
which corresponds to
\begin{eqnarray}
  P={g_sN_1-N\over N}=(g_s-1)~\!{N_1-N_2\over N}~\!,
\end{eqnarray}
ceases to have the simple interpretation of the system's
polarization.\footnote{Obviously, if the spontaneous ordering is approached
  in the usual thermodynamic limit by switching off an external magnetic
  field coupled to the fermion spin, it will be the largest spin projection
  in the direction of the field that will be singled out. In this case
  the magnetization will be given by $N\mu P/(g_s-1)$, where $\mu$
  is the magnetic moment associated with a single spin.}
In terms of the factors $x_1(P)$ and $x_2(P,g_s)$  given by
(\ref{eqn:x1Andx2Defs}) the curly bracket in the formula
(\ref{eqn:EnergyUpToSecondOrder}) takes the form
\begin{eqnarray}
  f_s(P)=x_1^5+(g_s-1)x_2^5+{20\over9\pi}~\!k_{\rm F}a_0~\!(g_s-1)~\!x_2^3
  \left[x_1^3+{1\over2}~\!(g_s-2)~\!x_2^3\right]\nonumber\\
  +~\!{2\over\pi^2}~\!(k_{\rm F}a_0)^2(g_s-1)\left[J_K(x_1,x_2)+
    {1\over2}~\!(g_s-2)~\!{44-8\ln2\over21}\right].\phantom{a}~\!
  \label{eqn:EnergyUpToSecondOrderReduced}
\end{eqnarray}

In the case of spin $1/2$ fermions the phase transition to the ordered
state at $k_{\rm F}a_0=\pi/2$ in the mean field approximation, equivalent
to retaining in (\ref{eqn:EnergyUpToSecondOrder}) only the first two
terms in the curly bracket, is known to be continuous. This results
however from a numerical coincidence which makes the second derivative
of $E^{\rm MF}_\Omega(P)$ with respect to $P$ to vanish at $P=0$ for
$k_{\rm F}a_0=\pi/2$, precisely when the nontrivial minimum first appears.
This does not happen if $s>1/2$ and in the mean field approximation the
phase transitions to the ordered states are in all these cases of
first order.\footnote{Taking the effects of the interaction in the
  first nontrivial order and restricting oneself to the configurations
  specified by (\ref{eqn:x1Andx2Defs}) one can repeat the steps taken
  in \cite{Kesio} and obtain the full low temperature profile of the transition
  to the ordered phase for any spin $s$. At zero magnetic field $H$,
  the equation
  \begin{eqnarray}
    g_s\mu H+{4g_s\over3\pi}~\!k_{\rm F}a_0~\!P
    =(g_s-1)\left\{(1+P)^{2/3}-\left(1-{P\over g_s-1}\right)^{2/3}
    \right.\phantom{aaaaaaaaaaaaaaaaaaaaaaa}\nonumber\\
    \left.
    -{\pi^2\over12}\left({k_{\rm B}T\over\varepsilon^{(s)}_{\rm F}}\right)^2
    \left[(1+P)^{-2/3}-\left(1-{P\over g_s-1}\right)^{-2/3}\right]
    +\dots\right\},\nonumber
  \end{eqnarray}
  which determines the parameter $P$ starts, for $s>1/2$ to have
  a nontrivial solution before the slopes at $P=0$ of its both sides
  equalize - this shows that the  transition is of first order.}
It can be also easily checked that the critical value of
$k_{\rm F}a_0$ at which it occurs decreases with the spin $s$ but rather
slowly - $(k_{\rm F}a_0)_{\rm cr}^{\rm MF}\approx1.43$ for $s=3/2$, 
$(k_{\rm F}a_0)_{\rm cr}^{\rm MF}\approx1.21$ for $s=7/2$ and
$(k_{\rm F}a_0)_{\rm cr}^{\rm MF}\approx1.14$ for $s=9/2$. 
In view of the qualitative explanation of the reasons for the transition
such a behaviour is easy to understand: the gas kinetic energy is linear
in the number $g_s$ of the possible spin projections while the interaction
depends on it (in the mean field approximation) quadratically. One can
also observe that only for $s=3/2$ (and, of course, $s=1/2$) is the
spontaneous polarization $P$ not maximal at $(k_{\rm F}a_0)_{\rm cr}^{\rm MF}$;
for all higher spins it becomes maximal ($P=g_s-1$) already at the transition
point.

If the corrections of order $(k_{\rm F}a_0)^2$ are included, i.e. the full
formula (\ref{eqn:EnergyUpToSecondOrderReduced}) is used, the transition
to the ordered state (at zero temperature) becomes of first order also for
$s={1\over2}$ (and retains, of course, this character for $s>{1\over2}$).
The relevant plots can be found in \cite{CHWO2} and in \cite{PECABO}
for $s={1\over2}$ and in \cite{PECABO} for $s\geq{1\over2}$.
The transition occurs now at
$(k_{\rm F}a_0)_{\rm cr}^{\rm2nd}\approx1.054$ for $s=1/2$,
$(k_{\rm F}a_0)_{\rm cr}^{\rm2nd}\approx0.954$ for $s=3/2$,  
$(k_{\rm F}a_0)_{\rm cr}^{\rm2nd}\approx0.840$ for $s=7/2$ and
$(k_{\rm F}a_0)_{\rm cr}^{\rm2nd}\approx0.804$ for $s=9/2$. Moreover, in this
approximation only for $s=1/2$ the polarization gradually approaches
the maximal value $P=1$ ($P\approx0.58$ exactly at the critical coupling
$k_{\rm F}a_0=(k_{\rm F}a_0)_{\rm cr}^{\rm2nd}$); in the remaining cases
it jumps immediately to the maximal possible value $P_{\rm max}=g_s-1$.
\begin{figure}
\centerline{\hbox{
\psfig{figure=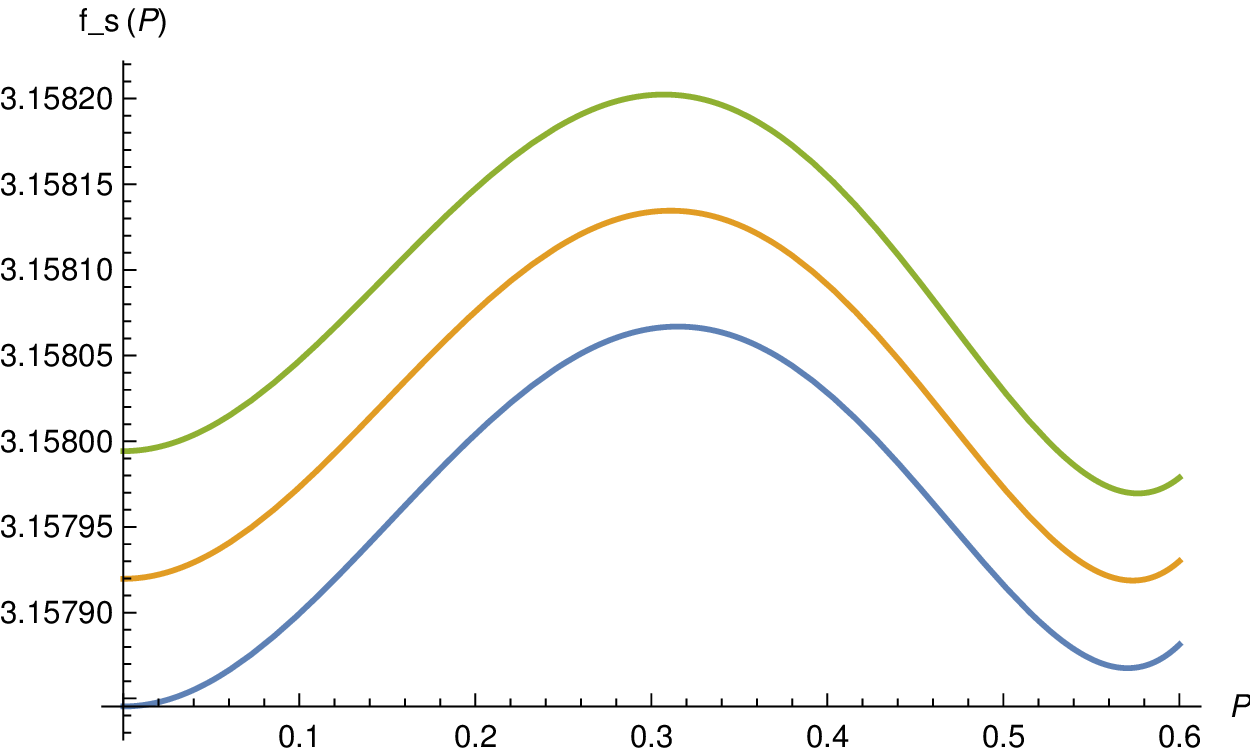,width=8.cm,height=5.0cm} 
\psfig{figure=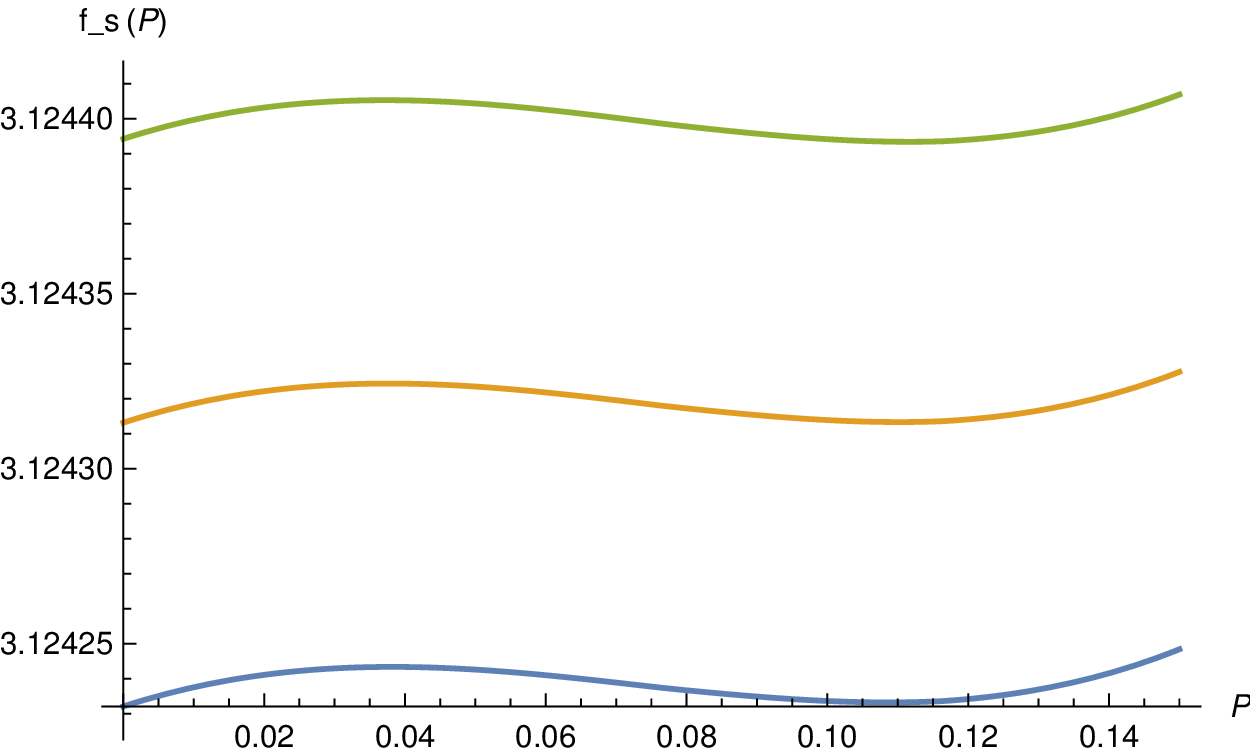,width=8.cm,height=5.0cm} 
}}
\caption{Plots of the ground state energy of the system of spin $1/2$
  fermions as a function of the order parameter $P$ in units of
  $(k_{\rm F}^3/6\pi^2)(3/5)(\hbar^2k_{\rm F}^2/2m_f)$, i.e. of the
  function $f_s(P)$. Left: without the order $(k_{\rm F}a_0)^3$ corrections - the
  lines from below correspond to $k_{\rm F}a_0=1.05404$ (blue), $1.05409$
  (red) and 1.105414 (green). Right: with the order $(k_{\rm F}a_0)^3$ corrections
  - the lines from below correspond to $k_{\rm F}a_0=0.99091$ (blue), $0.99096$
  (red) and 0.99101 (green). In both panels the middle line corresponds
  to the respective critical value of $k_{\rm F}a_0$.}
\label{fig:Egs1}
\end{figure}

\begin{figure}
\centerline{\hbox{
\psfig{figure=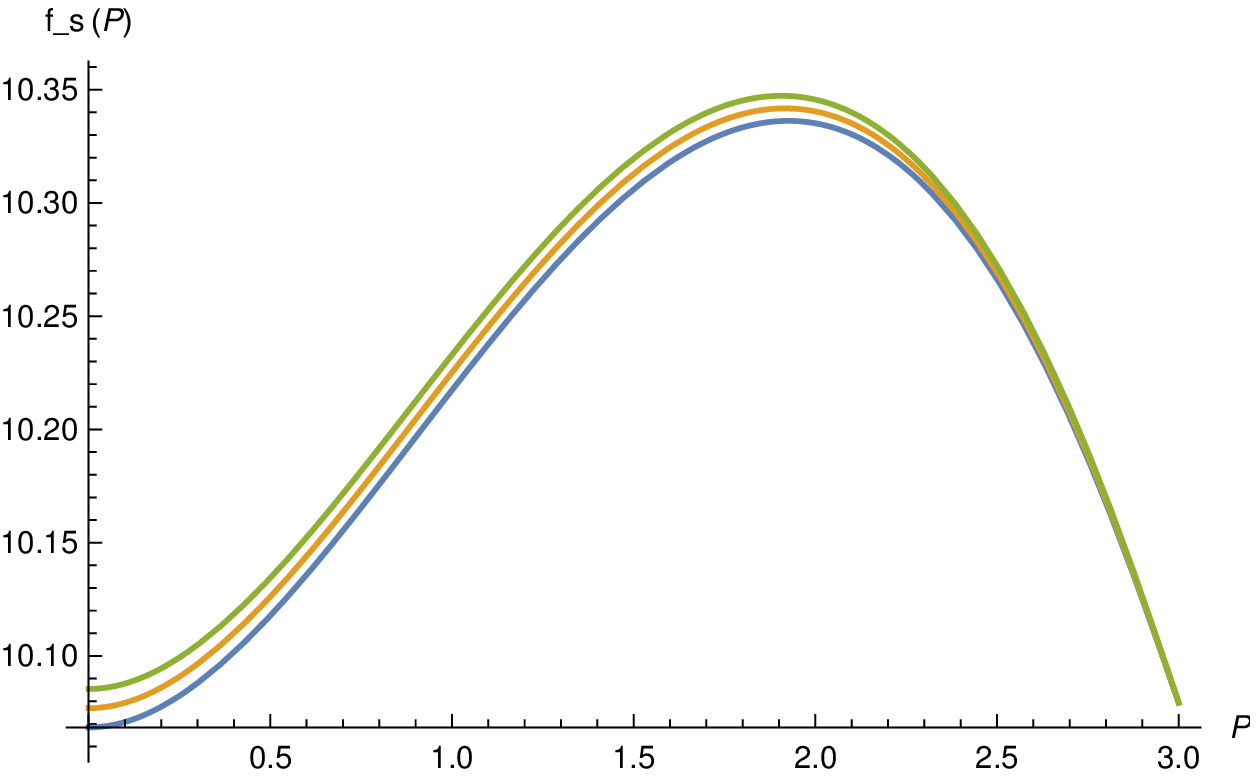,width=8.cm,height=5.0cm} 
\psfig{figure=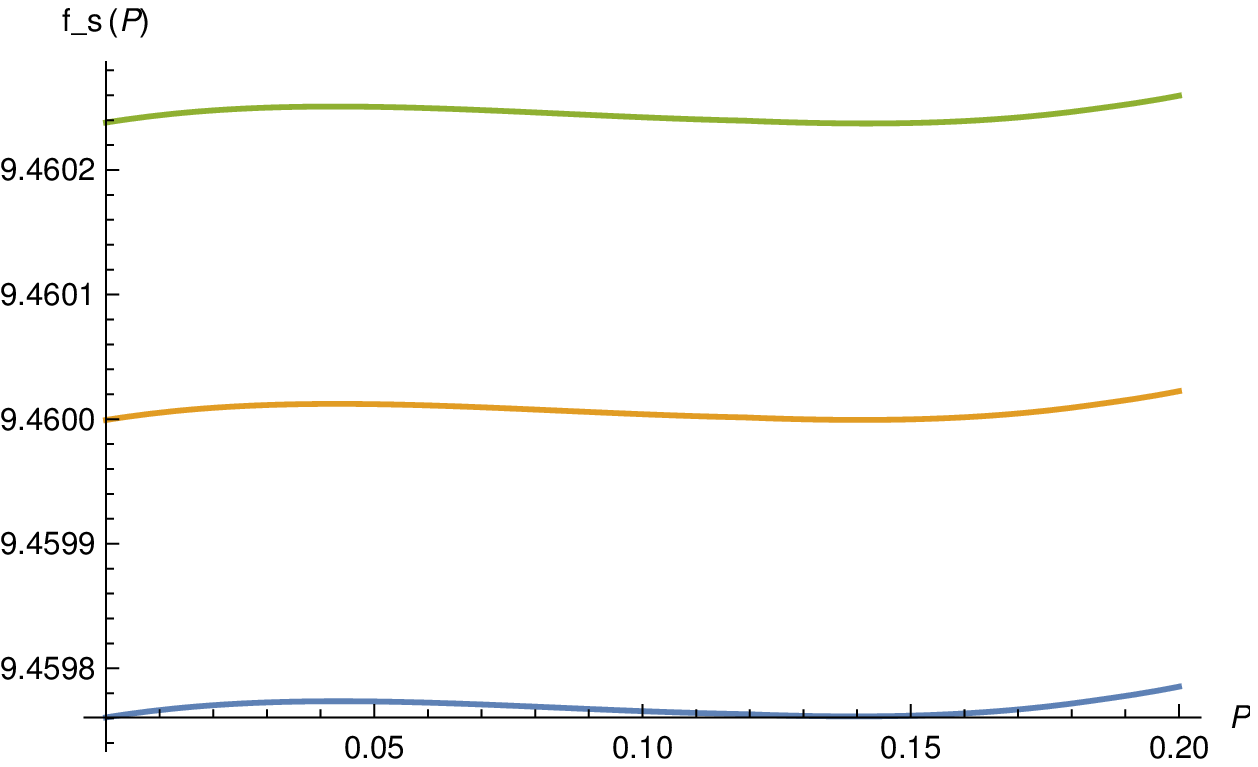,width=8.cm,height=5.0cm} 
}}
\caption{Plots of the ground state energy of the system of spin $3/2$
  fermions as a function of the order parameter $P$ in units of
  $(k_{\rm F}^3/6\pi^2)(3/5)(\hbar^2k_{\rm F}^2/2m_f)$, i.e. of the
  function $f_s(P)$. Left: without the order $(k_{\rm F}a_0)^3$ corrections
  - the lines from below correspond to $k_{\rm F}a_0=0.9532$ (blue), $0.9542$
  (red) and $0.9552$ (green). Right: with the order $(k_{\rm F}a_0)^3$ corrections
  - the lines from below correspond to $k_{\rm F}a_0=0.74188$ (blue), $0.74190$
  (red) and $0.74192$ (green). In both panels the middle line corresponds
  to the respective critical value of $k_{\rm F}a_0$.}
\label{fig:Egs4}
\end{figure}

Inclusion of the order $(k_{\rm F}a_0)^3$ corrections to the energy
density treated as a function of the order parameter $P$ amounts to adding
to the function $f_s(P)$ given by (\ref{eqn:EnergyUpToSecondOrderReduced})
the term
\begin{eqnarray}
  {160\over\pi^3}~\!(k_{\rm F}a_0)^3(g_s-1)\left[4G^{(1)}_{\rm fin}(x_2,x_1)
    +4G^{(2)}(x_2,x_1)
    -K^{(1)}(x_2,x_1)-K^{(2)}(x_2,x_1)\right.\phantom{aaaa}\nonumber\\
    +{1\over2}(g_s-2)\left(4G^{(1)}_{\rm fin}(x_2,x_2)+4G^{(2)}(x_2,x_2)
    -K^{(1)}(x_2,x_2)-K^{(2)}(x_2,x_2)\right)\phantom{aa}\\
    \left.+(g_s-2)\left(\tilde K^{(2)}(x_1;x_2,x_2)+2\tilde K^{(2)}(x_2;x_2,x_1)
    +(g_s-3)\tilde K^{(2)}(x_2;x_2,x_2)\right)\right].\nonumber
\end{eqnarray}
This has the following consequences. Firstly, the critical value of the
expansion parameter $k_{\rm F}a_0$ is further reduced: the transition occurs
now at $(k_{\rm F}a_0)_{\rm cr}^{\rm3rd}=0.99096$ for $s=1/2$ (at
$(k_{\rm F}a_0)_{\rm cr}^{\rm3rd}=0.74190$ and $0.49445$ for $s=3/2$ and $s=9/2$,
respectively). Secondly, the first order character of the transition, quite
clear without this correction, becomes now much less pronounced - while for
$s=1/2$ the height of the hill separating the minimum at $P=0$ from the one
at $P\neq0$ remains almost unchanged, greatly reduced is the nonzero value
of the order parameter right at the transition. This is clear from the
comparison of the two panels of Figure \ref{fig:Egs1} showing the shape of 
the function $f_s(P)$ for three values of $k_{\rm F}a_0$ close to the critical
one (corresponding to the approximation used): in the left one the order
$(k_{\rm F}a_0)^3$ corrections are not included, in the right one they are.
This aspect - the reduction of the the nonzero value of the order parameter
right at the transition - is even more dramatic in the case of
higher spins: without the third order corrections this value
was maximal $P=g_s-1$; with these corrections it is reduced to $\sim0.12$
(practically independent of the value of $s$)! Moreover, for $s>1/2$
also the height  of the hill separating the symmetry breaking minimum
from the symmetry preserving one at $P=0$ is reduced by roughly three
for $s=3/2$ up to four for $s=9/2$ orders of magnitude. In the case of spin
$3/2$ this is shown in Figure \ref{fig:Egs4}. The results for 
different values of the spin $s$ are  presented in Table \ref{t:results}.

These purely perturbative results strongly point towards the possibility
that the transition to the ordered state is, when all effects associated
with the interactions other than that proportional to the coupling $C_0$
in (\ref{eqn:Veff}) are ignored (which is equivalent to setting to zero all
the higher partial waves scattering lengths $a_\ell$ and all the effective
radii $r_\ell$), indeed of the continuous
type. Most probably with the inclusion of higher and higher corrections
its character becomes less and less first order and becomes truly continuous
when a resummation of the sort performed in \cite{He} is made but
becomes practically indistinguishable from such already at a finite
(not very high, as our computation shows) order of the ordinary perturbative
expansion.

\vskip0.1cm

Finally we can briefly discuss effects induced by the potential presence
of nonnegligibly small $s$-wave effective radius $r_0$ and/or $p$-wave
scattering length $a_1$. They are introduced through the two lower length
dimension operators in (\ref{eqn:Veff}) which are given explicitly by the
formulae (\ref{eqn:C2Interaction}) and (\ref{eqn:C2PrimOpExplicit}). In
the configuration of densities assumed in this Section these interactions
add to the function $f_s(P)$ the terms
\begin{eqnarray}
  {1\over6\pi}~\!k_{\rm F}^3a_0^2r_0~\!(g_s-1)\left[x_1^3x_2^3(x_1^2+x_2^2)
    +(g_s-2)~\!x_2^8\right]\phantom{aaaaaaaaa}~\nonumber\\
  +{2\over3\pi}~\!(k_{\rm F}a_1)^3\left[x_1^8+{1\over2}g_s(g_s-1)~\!x_2^8
    +{1\over2}(g_s-1)~\!x_1^3x_2^3(x_1^2+x_2^2)\right].
  \label{eqn:HigherOrderOpsContrib}
\end{eqnarray}
If the effects of these operators are subleading compared to the effects
caused by the order $k_F^3$ corrections of the operator proportional to $C_0$
(that is, when that $|r_0|, |a_1|\ll a_0$) they modify the general picture
described above only slightly. For instance in the case of  $s=1/2$ the
presence of $r_0\approx a_0/10$ decreases by less than one percent the
critical value of the parameter $k_{\rm F}a_0$, while $r_0\approx -a_0/10$
increases it by the same amount in agreements with the observation made in
\cite{He}. In both cases the character of the phase transition is not
appreciably changed. Somewhat surprising is, however, the observation that
for larger spin values this patern is reversed: negative $r_0$ decreases
slightly the critical value of $k_{\rm F}a_0$, while positive $r_0$ increases
it. The reason for this is that while for $s=1/2$ the first line of 
(\ref{eqn:HigherOrderOpsContrib}) is a function of $P$ monotonically
decreasing to zero (the contribution to the energy density of the interaction
proportional to $C_2$ is by the Pauli exclusion principle also bound to
vanish at $P=1$), for higher spins it a is slightly increasing function,
precisely in the range in which the new minimum of the energy density forms.

In general, the correction induced by the operator
(\ref{eqn:C2PrimOpExplicit})  is larger than the one induced by
(\ref{eqn:C2Interaction}). Moreover in the perturbation series for
$E_\Omega/V$ it is the first correction which is not bound to vanish at
$P=1$ by the Pauli exclusion principle. 
Nevertheless, if one assumes that $|a_1|$ is of the same order as $|r_0|$
and much smaller than $a_0$, its contribution is not larger than
that of (\ref{eqn:C2Interaction}) simply because it is proportional to
$a_1^3$.  In the configuration of the densities considered here the expression
in the second line of (\ref{eqn:HigherOrderOpsContrib}) is for all values of the
spin $s$ a monotonically increasing function. For this reason the
contribution increases the value of $(k_{\rm F}a_0)_{\rm cr}$ for $a_1>0$
and lowers if for $a_1<0$ by roughly one percent so long as
$|a_1|\simlt a_0/10$ without affecting significantly the character of the
transition. Thus, it seems likely, that if all the scattering lengths $a_\ell$,
$\ell=1,2,\dots$, all the radii $r_\ell$, etc. are subleading with respect to
$a_0$, the continuos character of the transition will emerge after
all corrections are taken into account (being, from the practical
point of view indistinguishable from continuos already starting from
some fixed order of the perturbative expansion).

\begin{table}
\begin{center}
$
\begin{array}{|c||c|c||c|c||c|c|}
\hline
s& (k_Fa_0)_{\rm cr}^{\rm MF} & P^{\rm MF}_{\rm cr}&(k_Fa_0)_{\rm cr}^{\rm 2nd}&P^{\rm 2nd}_{\rm cr}
&(k_Fa_0)_{\rm cr}^{\rm 3rd}&P^{\rm 3rd}_{\rm cr}\\
\hline
 1/2& \pi/2 & 0        & 1.0540& 0.58     & 0.99097& 0.11\\ 
 3/2& 1.4298& 2.78     & 0.9543& P_{\rm max}& 0.74190& 0.14\\ 
 5/2& 1.3016& P_{\rm max}& 0.8877& P_{\rm max}& 0.62560& 0.125\\ 
 7/2& 1.2118& P_{\rm max}& 0.8405& P_{\rm max}& 0.54976& 0.115\\ 
 9/2& 1.1439& P_{\rm max}& 0.8040& P_{\rm max}& 0.49445& 0.105\\ 
11/2& 1.0901& P_{\rm max}& 0.7753& P_{\rm max}& 0.45164& 0.10\\ 
\hline
\end{array}
$
\caption{Characteristics of the transitions to the ordered state at $T=0$
for different values of the spin $s$ of fermions.}
\label{t:results}
\end{center}
\end{table}

\section{Conclusions}

Computing the ground-state energy of a finite density system of fermions
interacting through a binary spin independent repulsive interaction is a
classic problem of many-body quantum mechanics. The modern effective field
theory approach greatly simplifies this task. In particular it allowed to
complete the computation of the fourth order (i.e. proportional to $k_{\rm F}^4$)
corrections in the case of spin $s$ fermions and equal densities of
different spin projections \cite{WeDrSch} and to extend to the third order the
computation of energy of spin $1/2$ fermions for an arbitrary value of the
system's polarization \cite{CHWO3}. Here we have extended the latter result
to the case of spin $s$ fermions and arbitrary densities of different spin
projections. The derived formulae allow to compute the ground state energy
semi-analytically exploiting only the built-in routines of the Mathematica
package.

We have used this result to discuss two issues. Firstly we have checked
numerically how the magnitude of the third order term included in the
resummation of an infinite subset of corrections done in \cite{He} compares
with the magnitudes of the rejected terms and found that it is not obviously
dominant. Secondly, we have discussed the impact the third order corrections
have on the characteristics of the system's transition to the ordered phase
at zero temperature. We have found that already the third order corrections
tend to erase the first order character of this transition independently
of the value $s$ of the spin of fermions. Although our observations are
made on the basis of the perturbative expansion used in the regime which
is probably beyond the domain of its applicability (the comparison of
the results of the Quantum Monte Carlo computations of the ground state
energy of the unpolarized gas of spin $1/2$ fermions with the perturbative
coputation reveals \cite{QMC10} that the expansion is reliable up to
$k_{\rm F}a_0\simlt0.6$; in the case of $s>1/2$ the value of the expansion
parameter
$k_{\rm F}a_0$ at which the transition to the ordered state occurs is smaller
but since at the same time the magnitude of the successive terms of the
perturbation series are increased by the growing powers of the factor $g_s$,
the limit of the reliability of the expansion probably also decreases with
$s$), they nevertheless seem to lend some support to the
claim made in \cite{He} that the transition is continuous rather than
first order. It seems that the picture which emerges is quite sensible:
since continuous transition are associated with fluctuations at all
length scales, at any finite order of the perturbative expansion such
a transition should look like a first order one (the continuous transition
in the system of spin $1/2$ fermions obtained in the mean field
approximation is from this perspective a mere numerical coincidence),
though its first order character may rapidly disappear with increased
order of the approximation (so that it quickly may become
indistinguishable from a continuous one from the practical point of view);
the truly continuous character of a transition can really be revealed
only by a (partial) resummation of of all orders contributions, of the sort
proposed in \cite{He}.
\vskip0.3cm

\noindent{\bf Acknowledgments.} We would like to thank Dr Krzysztof
Jachymski for a discussion.


\end{document}